\documentclass[journal]{IEEEtran}
\usepackage{cite}
\usepackage{amsmath,amssymb,amsfonts}
\usepackage{algorithmic}
\usepackage{graphicx}
\usepackage{color}
\usepackage{textcomp}
\usepackage[T1]{fontenc}
\usepackage{url}
\usepackage[font=footnotesize]{caption}
\usepackage{subfig}

\ifCLASSINFOpdf
\else
\fi

\ifCLASSINFOpdf
\else
\fi
\begin{document}

%


\title{Multi-Antenna Configuration with Reduced Passive Self-Interference for Full-Duplex Intelligent Transportation System}

%
%

\author{Jogesh Chandra Dash,~\IEEEmembership{Member,~IEEE,}
        Debdeep Sarkar,~\IEEEmembership{Member,~IEEE}
\thanks{J. C. Dash and D. Sarkar are with the Department
of Electrical Communication Engineering, Indian Institute of Science, Bangalore,
Karnataka, 560021, India e-mail: (jcdash92@gmail.com and debdeep@iisc.ac.in.)}
\thanks{}
\thanks{}}

\maketitle
\pagenumbering{gobble}
\begin{abstract}
In this paper, we propose a closely spaced multi-antenna system with \textit{passive} self-interference cancellation (\textit{p}-SIC) of $\approx 90$ dB between the transmitter and receiver antenna for full-duplex application. The \textit{p}-SIC is achieved by field confinement near individual antennae using shorted metallic vias and the application of U-shaped perturbation in the ground plane. The \textit{p}-SIC technique is initially implemented in a 1-Tx and 1-Rx antenna system and explained using transmission line-based theory. Further, it is extended to 1-Tx and 2-Rx configurations. Here the proposed full-duplex antenna system is designed at $5.9$ GHz ($5.855-5.925$ GHz, IEEE 802.11p / WAVE technology) intelligent transportation system (ITS) application band using a microstrip patch configuration. The individual antenna exhibits an impedance bandwidth of $93$ MHz ($5.850-5.944$ GHz), $5.63$ dBi gain at $5.9$ GHz operating frequency and X-pol level less than $20$ dB in the broad side direction. The proposed FD configuration exhibits $|S_{ij}|$ of less than $-50$ dB over the complete operating band and $\approx -90$ dB is achieved at the operating frequency between the Tx and Rx. Similarly, $|S_{ij}|$ of less the $-30$ dB is achieved between 2-Rx antennas for a three-element FD configuration. The design procedure of the proposed FD configuration is explained and verified using fabrication and measurement. An experimental demonstration of the self-interference channel and its suppression using the proposed \textit{p}-SIC technique is also provided. Further, to study the diversity performance of the proposed multi-antenna configuration, the MIMO performance metrics such as \textit{ECC} and \textit{CCL} are evaluated using simulation and measurement.
\end{abstract}

\begin{IEEEkeywords}
Full-Duplex, Microstrip Antenna, Self-Interference Cancellation, Intelligent Transportation System (ITS), V2X, MIMO,  Mutual Coupling Reduction.
\end{IEEEkeywords}

%
\IEEEpeerreviewmaketitle

\section{Introduction}
%
%
%
%
\IEEEPARstart{I}{ntelligent} transportation system (ITS) is the future of vehicular wireless communication as the advanced driving experience and active safety are now priorities of every autonomous industry for upcoming vehicles. To reduce the impact on environment and to improve the traffic safety intelligent transportation systems (ITS) for vehicular communication such as vehicle-to-vehicle (V2V), vehicle-to-infrastructure (V2I) or in-short vehicle-to-everything (V2X) communication is the study of great interest for industry as well as academia \cite{yf}. A simplified architecture of ITS showing V2X connectivity is shown Fig. 1. In addition, ITS caters various smart applications such as safety warnings, emergency alert, vehicle platooning, vehicle internet, etc \cite{je}.

\begin{figure}[]
        \centering
         \includegraphics[width=\columnwidth,height=15 cm,keepaspectratio]{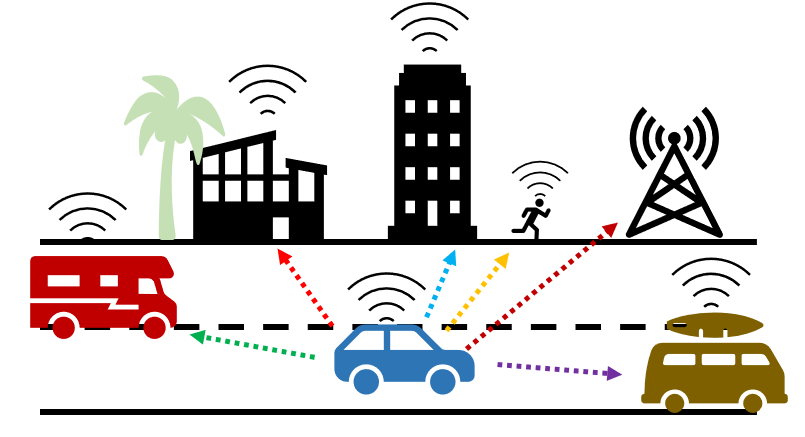}
         \caption{Simplified architecture of intelligent transportation system (ITS) showing V2X connectivity.}
         \label{fig1}
    \end{figure}
    
A report on legacy ITS and advanced ITS is provided in \cite{itu}. Accordingly, ITS is termed as cooperative-ITS (C-ITS) in Europe and some Asia-Pacific regions which work at the 5.855-5.925 GHz band. Similarly, in the USA and some other countries in North America dedicated short-range communication (DSRC) serves under ITS at the operating band of 5.850-5.925 GHz. Apart from C-ITS and DSRC, some other vehicular communication technologies such as V2X-WAVE (wireless access in the vehicular environment) and V2X- ITS connect come under ITS based on 5.9 GHz IEEE 802.11p standard \cite{itu}.

The ITS application scenario, in contrast to conventional mobile communication, requires an antenna system enabling ultra-high reliability and low latency communication \cite{cb}. There are certain single antenna configurations for ITS application available in literature \cite{gz} - \cite{ol}. In \cite{gz}, a low profile integrated microstrip antenna for dual-band application is proposed. A stacked slot antenna structure with additional slots on the top metallic part, floating metal layer, and reflector metal with additional dielectric layers is proposed in \cite{ti}, which is quite bulky and may face integration issues in a vehicle. Similarly, a single left-hand circularly polarised (LHCP) antenna having $40 \times 40$ mm$^2$ size is proposed in \cite{ol}. However, future autonomous vehicles are expected to have a multi-antenna configuration, with 5G connectivity, installed at different parts of the vehicle to provide $360^0$ field-of-view \cite{sr}, where a single antenna configuration is no longer a suitable solution for ITS application to cater the demand of ultra-high reliability and low latency.

Multi-antenna configuration based on full-duplex (FD) technology would be the best possible solution for ITS application considering the future requirements. Full-duplex topology concurrently uses time/frequency slots both for transmission and reception. This can be integrated with the 5G system over the conventional half-duplex (HD) system, which operates either frequency-division duplex (FDD) or time-division duplex (TDD), to improve the system efficiency \cite{ke}. A schematic of the HD and FD communication system is provided in Fig. 2 for better visualization. As far as ITS application is concerned, the simultaneous transmission and reception capability of FD system theoretically double the spectral efficiency of HD system \cite{hv}, which eventually reduces latency \cite{fa}. In addition, the FD technology improves throughput \cite{fa} and provides high-speed ultra-reliable V2X link in vehicular communication \cite{cc}. Moreover, FD technology is one of the candidate technologies for future 6G services which facilitates enhanced capabilities than 5G counterpart (see Table 1)) \cite{6g} and can efficiently serve the ITS purpose. However, the key challenge in an FD system is the issue of self-interference due to the infliction of high-power local transmit a signal upon its own receiver, which leads to device saturation and failure in receiving actual signal of interest \cite{ji}, \cite{jc}. To combat SI and make the FD system applicable in practice the high isolation of more than $100$ dB \cite{as} (however it may vary based on the type of wireless system \cite{hz}) between the transmitter and receiver unit is highly desirable. In general, such high isolation is achieved jointly by \textit{passive(p)} and \textit{active(a)} SI-cancellation (SIC) techniques, where \textit{p}-SIC deals with the antenna/propagation domain SIC and \textit{a}-SIC deals with rf/analog and digital domain SIC \cite{ke}, \cite{kg}. However, analog-SIC is not sufficient to provide enough isolation to restrict the saturation of active components such as LNA (low noise amplifier) and ADC (analog-to-digital converter) \cite{ab}. On the contrary, the antenna is the very first component in the rf front end which experiences all external stimuli and activates the rest of the rf component in the chain. Therefore, to reduce the system cost and burden on \textit{a}-SIC stage, \textit{p}-SIC technique (antenna isolation) is very much essential in an FD communication system.

	\begin{figure}[]
        \centering
         \includegraphics[width=\columnwidth,height=15cm,keepaspectratio]{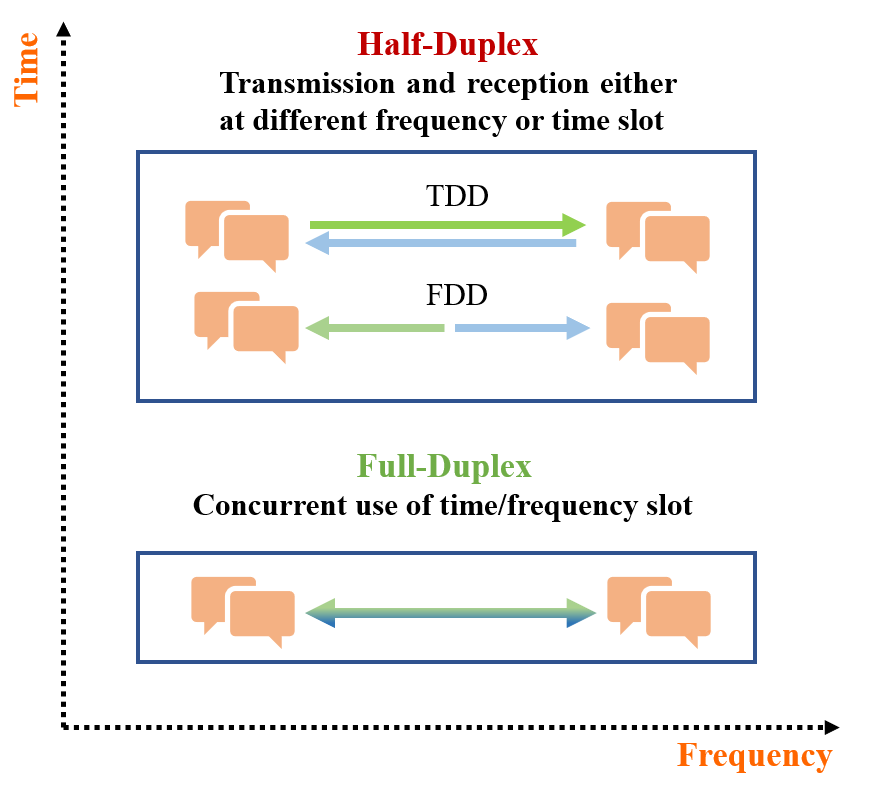}
         \caption{Schematic representation of HD and FD communication system.}
         \label{fig2}
    \end{figure}

There are various techniques have been reported to improve the inter-element isolation between the antenna elements, but either only in transmitting or receiving mode i.e in the HD scenario. For example, \cite{ks} provides an isolation improvement of $26.2$ dB using a parallel coupled line resonator between two microstrip antennas, isolation improvement of $22.3$ dB is achieved in \cite{yc} between two microstrip antennas using polarisation conversion isolator. Similarly, several other techniques such as defected ground structure \cite{hx}, split-ring resonator \cite{ds}, frequency selective surface \cite{th} etc. result in improved isolation around $15$ dB are reported. It shows that due to either only transmission or reception capability of HD system there is no issue of SI here, where it can operate smoothly having limited antenna isolation. However, this is not the scenario in the FD communication system.
There are various categories of SIC in FD antenna configuration available in the literature. Polarization multiplexing with hybrid-coupler-based SIC 
techniques is exploited in \cite{hn1} - \cite{hn3} and \cite{mv} to achieve very high inter-port isolation in expense of large size and stacked structure respectively. Similarly, a dual-polarized multi-layered antenna configuration with an additional feeding network and another dual polarized horn antenna configuration with high inter-port isolation for FD communication are proposed in \cite{ym} and \cite{fc} respective. However, due to their multi-layered and bulky structure, these are not suitable for integration in compact modules. A dual-polarized rectangular cavity-backed antenna with high inter-port isolation is described in \cite{ma}. The antenna has a 3D volumetric structure with very complex design and integration challenges. Co-linearly polarized compact size microstrip antenna with fence-strip resonator (FSR) for good isolation between transmitter and receiver port, a stacked structure FD antenna with the combination of a microstrip and surface integrated waveguide cavity backed slot antenna with cavity-patch coupling (CPC) path is added for cancelling the slot-patch coupling (SPC) by using four probe and  surface integrated wave guide based FD antenna with SIC using metallic vias are proposed in \cite{yh}, \cite{wz} and \cite{ai} respectively. However, the techniques proposed in \cite{yh} - \cite{ai} provides SIC between $30-40$ dB, which is quite limited for FD communication. In our earlier work, a very closely spaced planer FD antenna configuration having $90$ dB inter-port isolation, at the operating frequency, between the antenna elements is proposed in \cite{jc}.

\begin{table}[]
\centering
\caption{Performance Comparison between 5G and 6G \cite{6g}}
\begin{tabular}{ccc}
\hline
\textbf{Performance parameter} & \textbf{5G} & \textbf{6G} \\ \hline
Energy Efficiency & $1x$ & $2x$ \\
Spectral Efficiency & $1x$ & $2x$ \\
Air Latency (ms) & $1$ & $0.1$ \\
Connection Density (devices/km\textasciicircum{}2) & $10^{6}$ & $10^{7}$ \\
Reliability & $10^{-6}$ & $10^{-7}$ \\
Peak Data Rate (Gbps) & $20$ & $1000$ \\
User Experience Data Rate (Gbps) & $0.1$ & $1$ \\ \hline
\end{tabular}
\end{table}

Therefore, considering the application scenario and requirement for FD communication system in ITS, in this paper, we propose a very closely spaced planer easy to integrate FD antenna systems with a very high \textit{p}-SIC. This work builds upon work earlier work \cite{awpl} by: (i) a detailed explanation for 2-element FD-antenna (1-Tx and 1-Rx) design steps and working mechanism of proposed \textit{p}-SIC technique, (ii) extension of 2-element FD-antenna to 3-element antenna system having 1-Tx and 2-Rx (or 2-Tx and 1-Rx) configuration followed by fabrication, measurement and study of multi-antenna (MIMO) performance, and (iii) experimental demonstration of the real-time self-interference channel and its suppression using proposed \textit{p}-SIC technique. To the best of the authors’ knowledge, a compact FD antenna configuration having more than $85$ dB of  \textit{p}-SIC and antenna performance parameters such as a gain of $5.6$ dB, a cross-polar level less than $-20$ dB with good MIMO diversity for ITS application is not available in the open literature.  

\section{ Design and  \textit{p}-SIC Mechanism of two element (1-Tx and 1-Rx) Full Duplex Antenna System}
The 3D perspective view of the proposed two port (1-Tx and 1-Rx) FD antenna is shown in Fig. 3. The antenna is designed on the RTD 5880 dielectric substrate having dielectric constant $\epsilon_r = 2.2$, dielectric loss tangent $tan\delta = 0.0009$ and $1.6$ mm thickness ($h$). The overall substrate dimension is $L_g \times W_g (38$ mm $\times 80$ mm). In this section, the design stages of the proposed FD structure with the \textit{p}-SIC (antenna isolation) mechanism are discussed. For every design stage the corresponding E- and H- field distribution along the isolation parameter ($S_{ij}$, where $i \neq j$) are utilised to illustrate the operating mechanism. 


 
	\begin{figure}[]
        \centering
         \includegraphics[width=\columnwidth,height=15 cm,keepaspectratio]{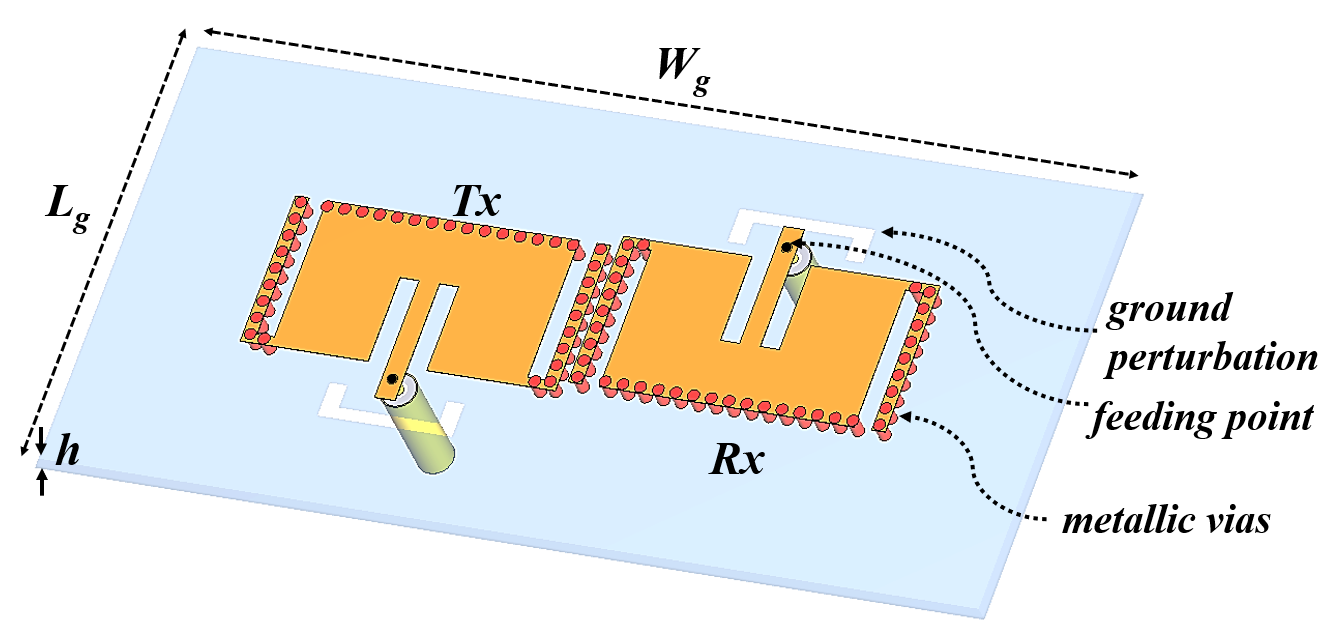}
         \caption{3D schematic representation of proposed 2-element FD antenna.}
         \label{fig3}
    \end{figure}

\subsection {Design Stage-1}
Initially, two element inset-fed microstrip patch antennas are placed in a closed proximity ($g$) with $180^0$ orientation representing Tx and Rx antenna for FD communication as shown in Fig. 4. The inset-fed structure is powered through a co-axial probe. The dimensions (see Fig. 4) of physical notch width ($g_1$), width of microstrip feed line ($w_f$) and the inset depth ($l_1$) are considered to maintain the individual antenna input impedance at the operating frequency which can be approximately computed using the following equations \cite{li}, \cite{balani}

\begin{equation}
   R_{in} = R_{in}(l_1 = 0).cos^4 \left(\frac{\pi l_1}{w}\right) 
\end{equation}
\begin{equation}
  R_{in}(l_1 = 0) = \frac{1}{2G} = \frac{1}{2}.\left(\frac{w}{120\lambda_0} \left[1 -\frac{(k_0h)^2}{24} \right] \right)^{-1} 
\end{equation}

where $R_{in}$ and $R_{in}(l_1 = 0)$ are the input impedance of the patch at $l_1$ and zero inset depth respectively. $G$ is the radiating edge conductance of the patch and $h$ denotes the substrate thickness. $k_0$ and $\lambda_0$ represent the free space wave-number and wavelength respectively.

The individual antenna element operates in $TM_{01}$ fundamental mode at $5.9$ GHz operating frequency having $3.38\%$ $-10$ dB impedance bandwidth ($\%$IBW)(See Fig. 5). Fig. 5 (a) and (b) depict the E- and H-field distribution showing the strong inter-element coupling between the antenna elements, which can also be observed from the S-parameter curve in Fig. 5(c) having very high value of $|S_{21}| = |S_{12}| = -11.8$ dB. This strong coupling is due to the close inter-element spacing and identical polarization of the Tx and Rx antennas.

   \begin{figure}[]
        \centering
         \includegraphics[width=\columnwidth,height=4 cm,keepaspectratio]{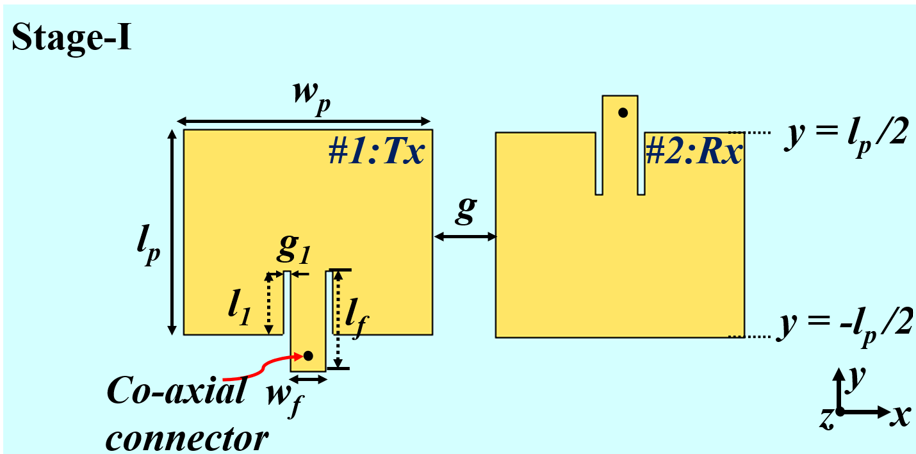}
         \caption{Schematic of two inset-fed patch antenna designed at 5.9 GHz operating frequency. Dimension (in mm): $l_{p} = 16.2, w_{p} = 21.6, l_f = 8, w_f = 2.75, l_1 = 5, g_1 = 0.625, g = 6.$}
         \label{fig3}
    \end{figure}
    
    \begin{figure}[]
        \centering
         \includegraphics[width=\columnwidth,height=5 cm,keepaspectratio]{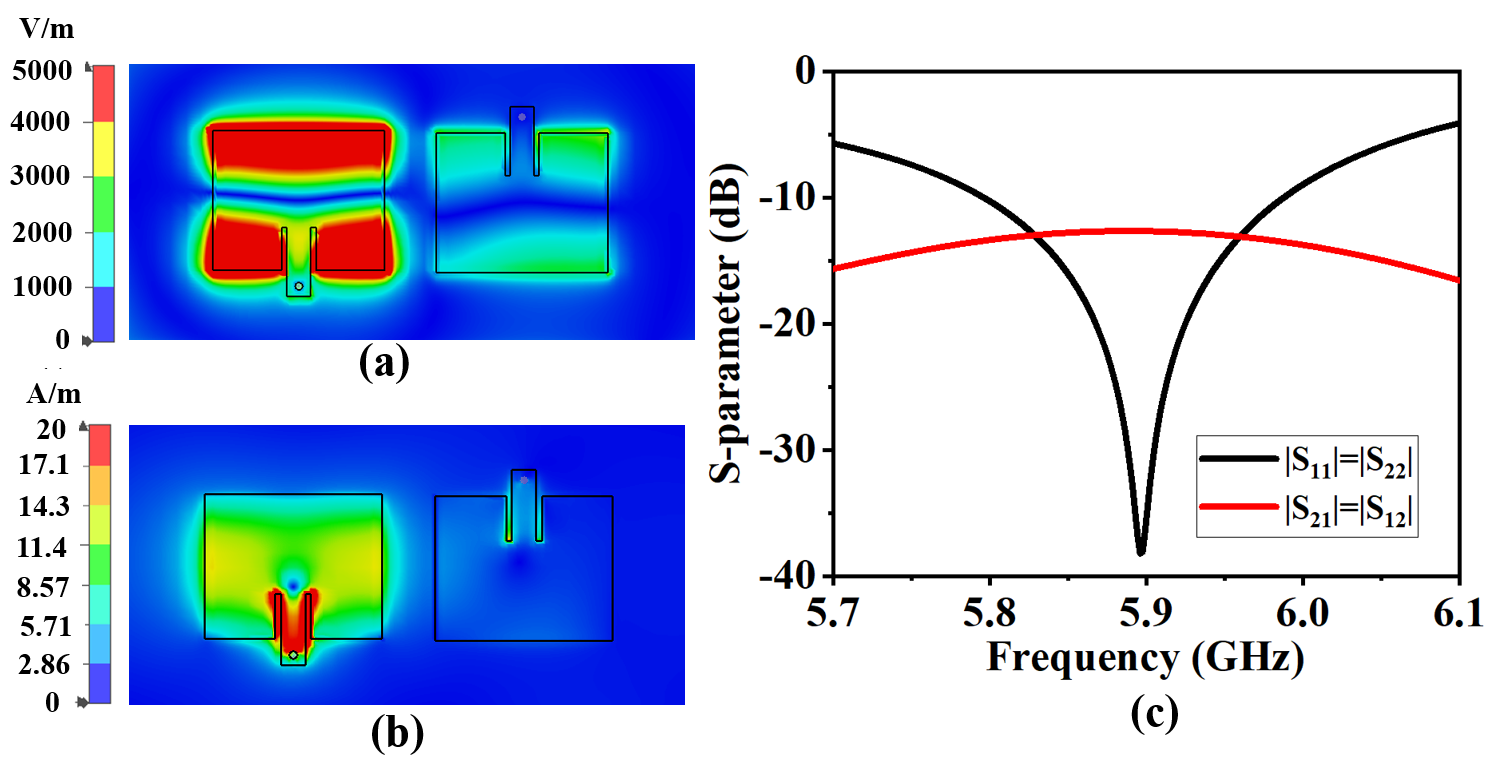}
         \caption{(a) E-field, (b) H-field distribution at 5.9 GHz (Port-1 is excited while terminating port-2 to matched load), and (c) Frequency variation of S-parameters of design Stage-1 as shown in Fig. 4.}
         \label{fig3}
    \end{figure}
  \subsection {Design Stage-II}  
Fig. 6 shows the schematic of antenna design at stage-II. In this design stage, initially two shorting vias are placed at two corners of Tx (@$-l_p/2$, see Fig. 4) and RX (@$l_p/2$, see Fig. 4) antenna elements. Further, a series of shorting vias are placed along at $g_3$ distance away from the non-radiating edges of the antennas along the $y$ direction, see Fig. 6. Here, the shorting via diameter ($d$) and distance between two consecutive vias ($p$) are $1$ mm and $1.5$ mm respectively. The objective of this design stage is reduce the high field coupling zone in the antenna elements and to limit the power flow between the Tx and Rx antenna by restricting the surface wave. It is evident from Fig. 5 (a) that the E-field coupling intensity has reduced compared to stage-I, which is also reflected in S-parameter response (see Fig. 7(c).  The $|S_{21}|$ parameter is reduced to $-19$ dB in this stage. The stage-II design exhibits no adverse effect to antenna polarization and operating frequency though it reduces the antenna $\%$IBW to $1.6$. 

\subsection {Design Stage-III}  
In this stage, to reduced the H-field coupling present in stage-II (see Fig. 7(b)) and to further reduce the E-field coupling, a series of shorting vias are placed at E-field minima positions (see Fig. 7(a)) of both Tx and Rx antenna along $x$-direction as shown in Fig. 8. The shorting pin dimension and the relation gap between the pins are kept identical as in stage-II. The loading of shorting vias along width of the patch at the zero potential location reduces the antenna size and made antenna operate at $TM_{0,1/2}$ mode. It is clearly visible in the Fig. 9 (a) and (b) that the changes made in this design stage reduces the field coupling compared to stage-II (see Fig. 7). Similarly, the $|S_{21}|$ response reduces to  $-31$ dB without affecting antenna impedance matching and bandwidth. The isolation achieved in this stage is suitable for HD applications, however this is not desirable for a FD systems.              

      \begin{figure}[]
        \centering
         \includegraphics[width=\columnwidth,height=4cm,keepaspectratio]{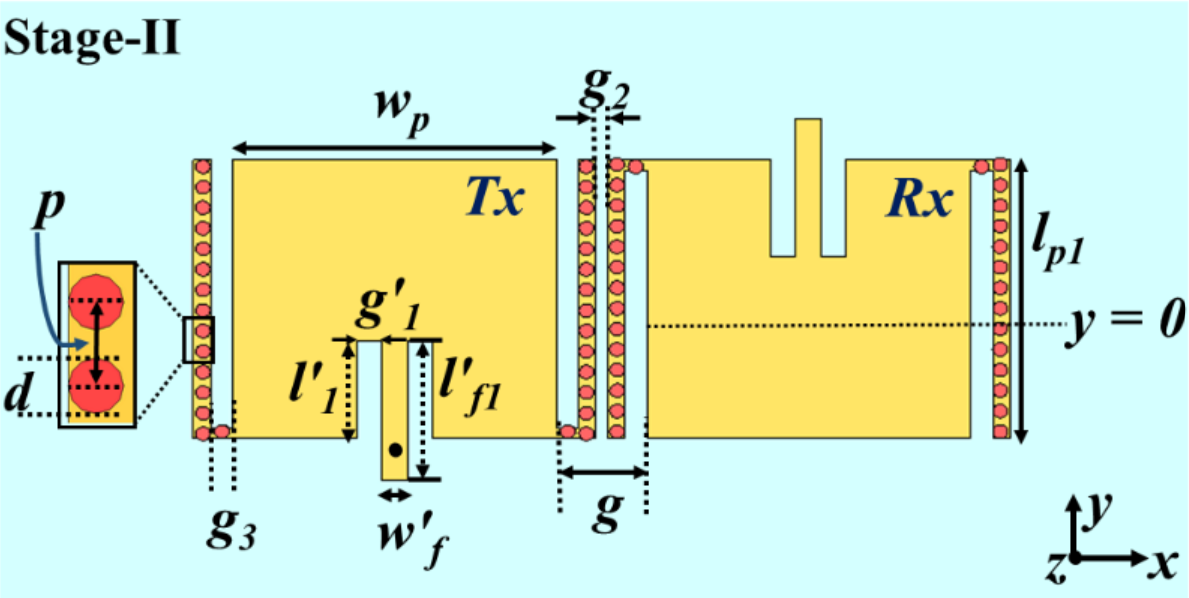}
         \caption{Schematic of two inset-fed patch antenna designed at 5.9 GHz operating frequency. Dimension (in mm): $l_{p1}=20.52$, $w_{p}=21.6$, $l'_{1}=7.1$, $g'_{1}=1.625$, $w'_{f}=1.75$, $l'_{f1}=10.1$, $g_{2}=0.75$, $g_{3}=1.5$, $g=6$.}
         \label{fig3}
    \end{figure}
  
  \begin{figure}[]
        \centering
         \includegraphics[width=\columnwidth,height=15 cm,keepaspectratio]{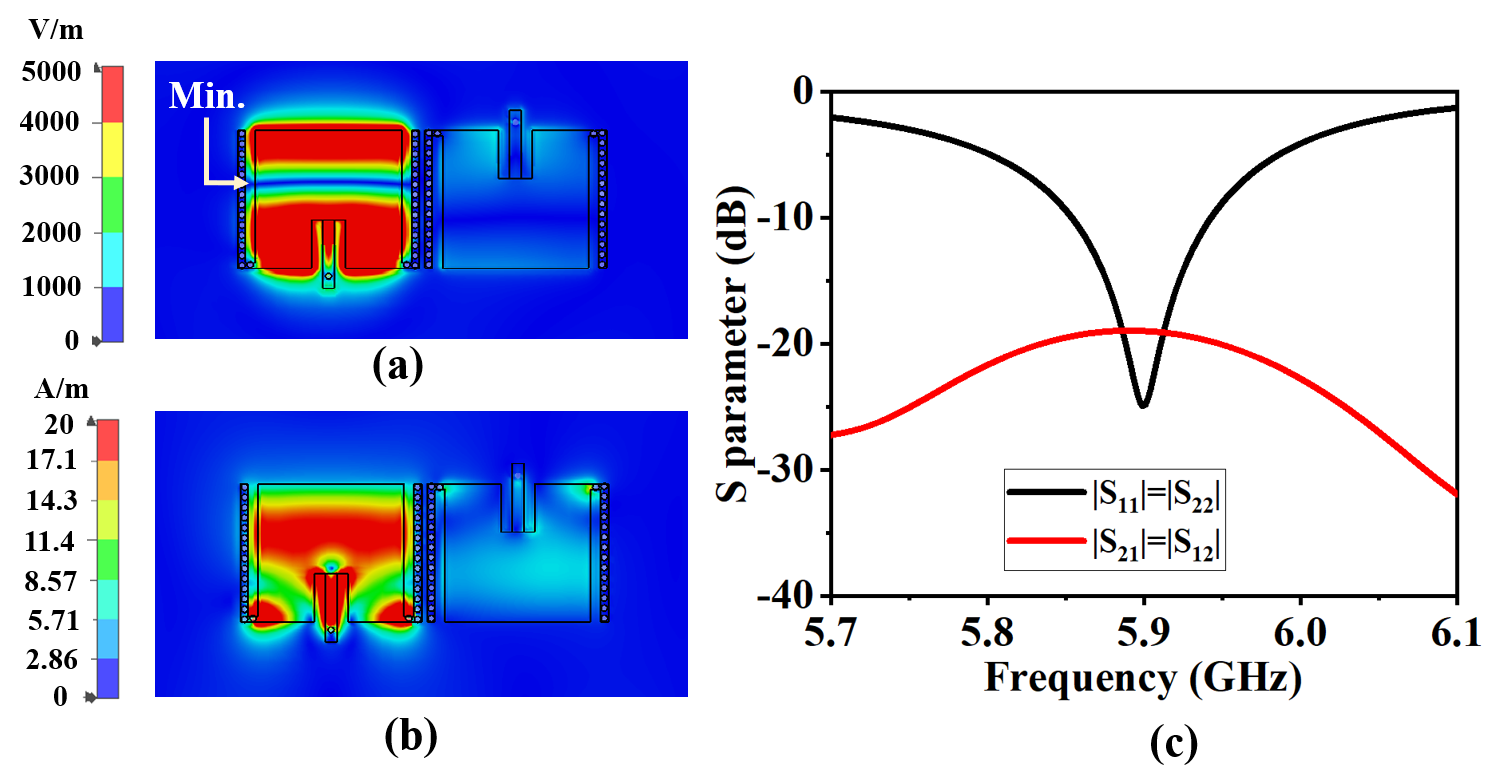}
         \caption{(a) E-field, (b) H-field distribution at 5.9 GHz (Port-1 is excited while terminating port-2 to matched load), and (c) Frequency variation of S-parameters of design Stage-1.5 as shown in Fig. 6.}
         \label{fig3}
    \end{figure}

       \begin{figure}[]
        \centering
         \includegraphics[width=\columnwidth,height=4 cm,keepaspectratio]{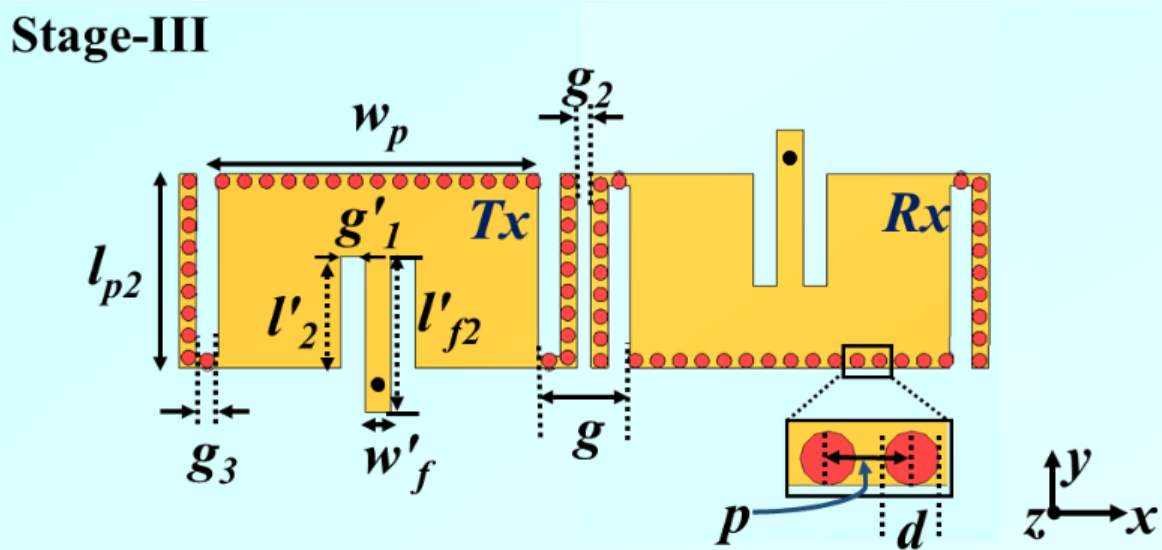}
         \caption{Schematic of two inset-fed patch antenna designed at 5.9 GHz operating frequency. Dimension (in mm): $l_{p2}=13.2$, $w_{p}=21.6$, $l'_{2}=7.6$, $g'_{1}=1.625$, $w'_{f}=1.75$, $l'_{f2}=10.6$, $g_{2}=0.75$, $g_{3}=1.5$, $g=6$, $d=1$, $p=1.5$.}
         \label{fig3}
    \end{figure}
    
    \begin{figure}[]
        \centering
         \includegraphics[width=\columnwidth,height=5 cm,keepaspectratio]{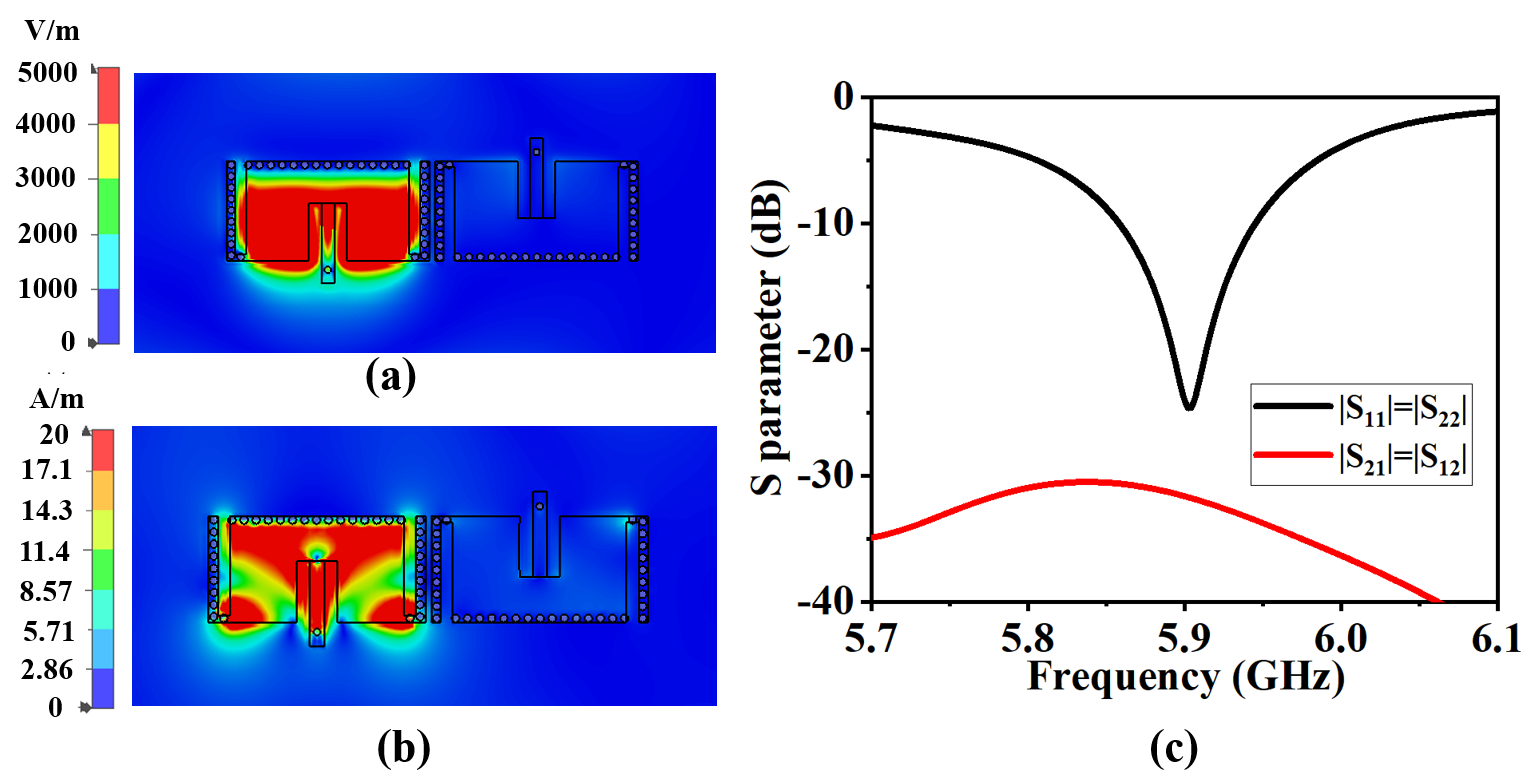}
         \caption{(a) E-field, (b) H-field distribution at 5.9 GHz (Port-1 is excited while terminating port-2 to matched load), and (c) Frequency variation of S-parameters of design Stage-1 as shown in Fig. 4.}
         \label{fig3}
    \end{figure}
    
\subsection {Design Stage-IV} 
In this stage, two U-shaped perturbations are incorporated in the ground plane near the feed locations of both Tx and Rx to achieve the desired \textit{p}-SIC (antenna isolation) between the Tx and Rx antennas. The schematic of the final FD antenna configuration is depicted in Fig. 10. The proposed antenna arrangement with the ground plane perturbation provides band-stop characteristics, which can easily be explained using the transmission line analogy.

     \begin{figure}[]
        \centering
         \includegraphics[width=\columnwidth,height=15 cm,keepaspectratio]{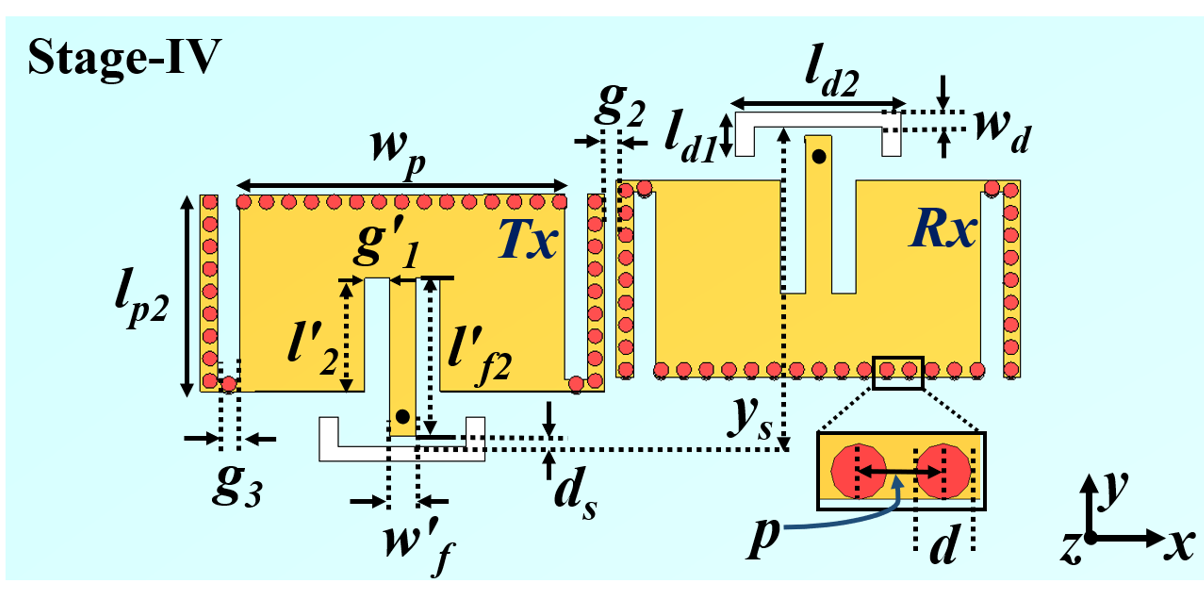}
         \caption{Schematic of two inset-fed patch antenna designed at 5.9 GHz operating frequency. Dimension (in mm): $l_{d1}= 3$, $l_{d2}= 11.5$, $w_{d}= 1$, $d_{s}=0.5$, $y_{s}=21.15$.}
         \label{fig3}
    \end{figure}

         \begin{figure}[]
        \centering
         \includegraphics[width=\columnwidth,height=15 cm,keepaspectratio]{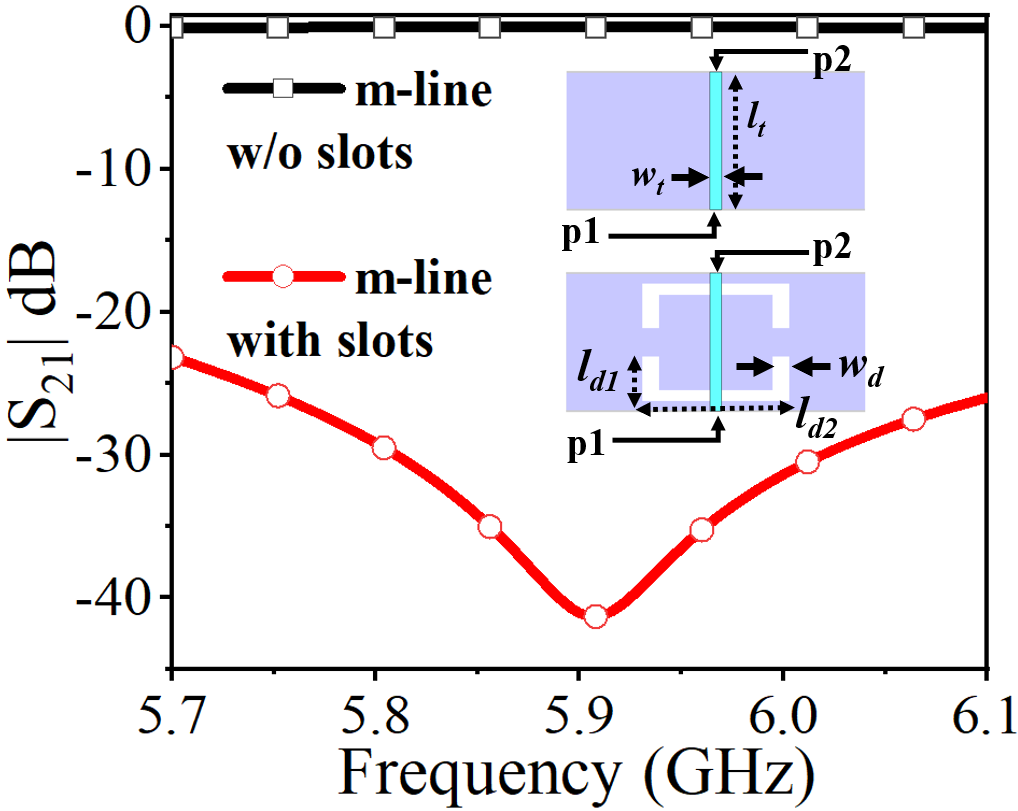}
         \caption{Variation of $S21$ with frequency for the microstriop transmission line (m-line) analogy of the proposed two port FD antenna as showing in Fig. 10 without and with DGS.(Dimensions in mm: $l_{t}= 12.4, w_{t}=1.05$), $l_{d1}= 4, l_{d2}= 13.22, w_{d}=1$).}
         \label{fig3}
    \end{figure}
    
    \begin{itemize}
    \item \textit{\textit{p}-SIC (Antenna Isolation) Mechanism using Transmission Analogy}: A microstrip transmission line is designed at $5.9$ GHz operating frequency, as shown in Fig. 11 (inset), on a RTD 5880 substrate to best describe the proposed antenna configuration. $l_t$, $w_t$, p1 and p2  are the length, width and the two ports of the transmission line respectively. $|S_{21}|= -0.05$ dB response of Fig. 11 for the transmission line without slots shows a passband response centered at $5.9$ GHz operating frequency. Further, two U-shaped perturbations are incorporated in the ground plane (i.e. defected ground structure (DGS)) of the transmission line near p1 and p2, as shown in Fig. 11 (inset). Similarly, the $|S_{21}|= -41.28$ dB response of the two DGSs loaded transmission line showing the stop-band response ($|S_{21}|=|S_{21}|$) at 5.9 GHz operating frequency is shown in Fig. 11. The design guidelines and equivalent circuit model of the DGS can be found from \cite{da}, where the length ($2l_{d1}+l_{d2}$) and width ($w_{d}$) of the DGS defines the resonant frequency which can be represented using an $LC$ equivalent circuit.
 \end{itemize}

The analysis can be re-direct to the stage-IV response as shown in Fig. 12. Fig. 12 (a) and (b) show the E- and H-field distributions of the final FD antenna configuration shown in Fig. 10, which depict approximately no field coupling between the Tx and Rx antenna element. This can also be observed in the S-parameter curve in Fig. 12 (c), where the isolation ($|S_{21}|$) is $87.08$ dB at the $5.9$ GHz operating frequency and more the $50$ dB over the operating band ($5.85-5.944$ GHz), which is very high compared to earlier stage-I and suitable for FD application. Dimensions of DGS, $g_2$ and $y_s$ are obtained using parametric study, as described in \cite{awpl} to get the best possible \textit{p}-SIC between the Tx and Rx antennas.

\begin{itemize}
\item \textit{Effect of Ground Plane on Proposed FD Antenna Performance}: The parametric studies on the antenna ground plane dimensions ($L_g$ and $W_g$) are conducted as shown in Fig. 13 (a) and (b). It can be observed in Fig. 13 (a) and (b) that isolation between the antenna elements varies with the variation of the antenna ground plane. The variation in the ground plane varies the electrical path length of the current distribution between the slots and eventually changes the effective inductive reactance introduced by the slots at the operating frequency \cite{cs}, thereby changing the isolation between antenna elements. However, the antenna $|S_{11}|$ parameter is unaffected (see Fig. 13(a) and (b)) as the ground plane variation does not affect the current distribution in the radiating patch. The stated design parameters for the proposed antenna are chosen for the ground plane size of $L_g \times W_g=38$ mm$ \times 80$ mm to adequately confined the microstrip antenna fringing field and to get the best possible antenna performance.

\end{itemize}

  \begin{figure}[]
        \centering
         \includegraphics[width=\columnwidth,height=15 cm,keepaspectratio]{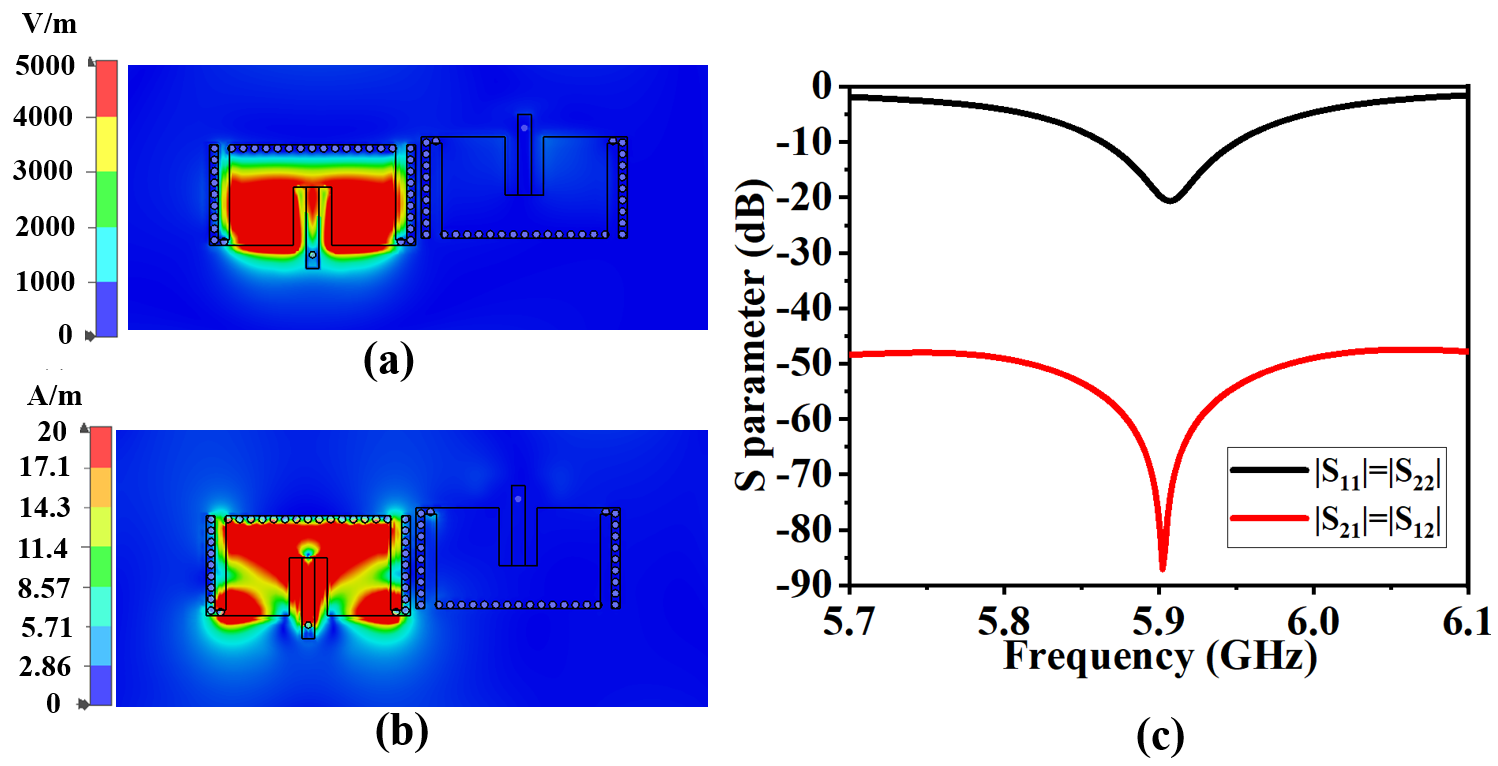}
         \caption{(a) E-field, (b) H-field distribution at 5.9 GHz (Port-1 is excited while terminating port-2 to matched load), and (c) Frequency variation of S-parameters of design Stage-2 as shown in Fig. 8.}
         \label{fig3}
    \end{figure}

\begin{figure}%
\centering
\subfloat[]{\includegraphics[width=0.8\columnwidth]{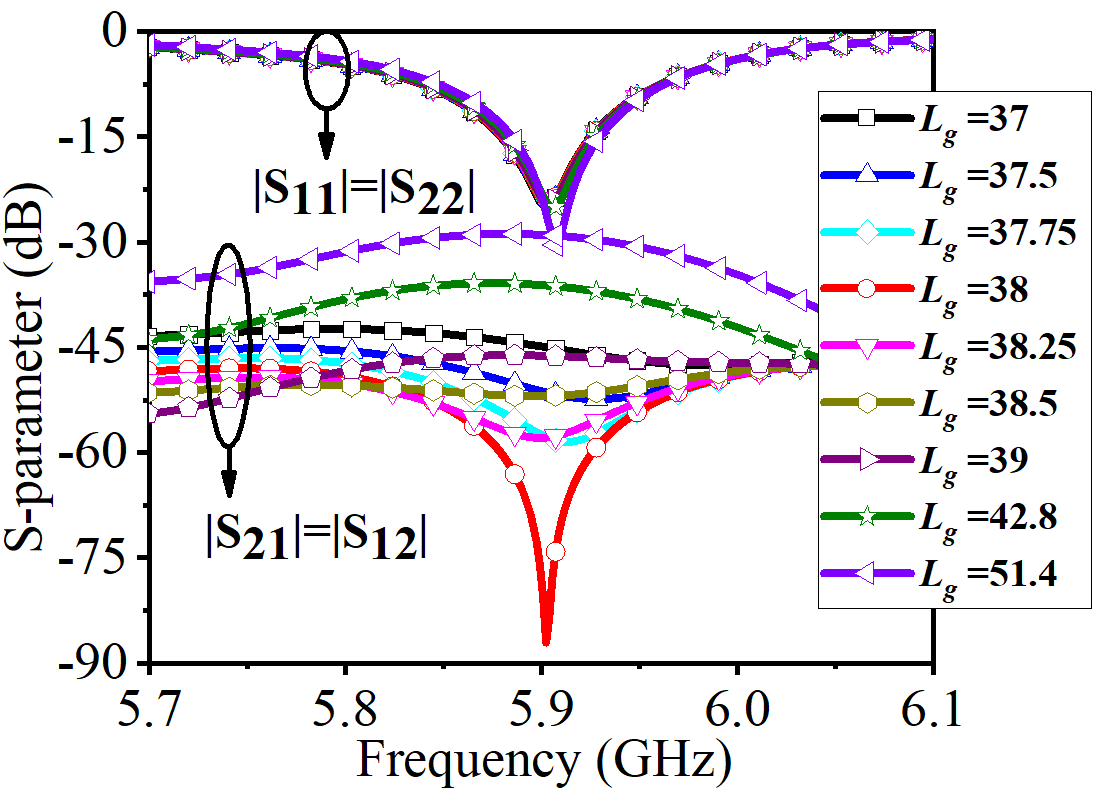}}%
\hspace{\fill}
\subfloat[]{\includegraphics[width=0.8\columnwidth]{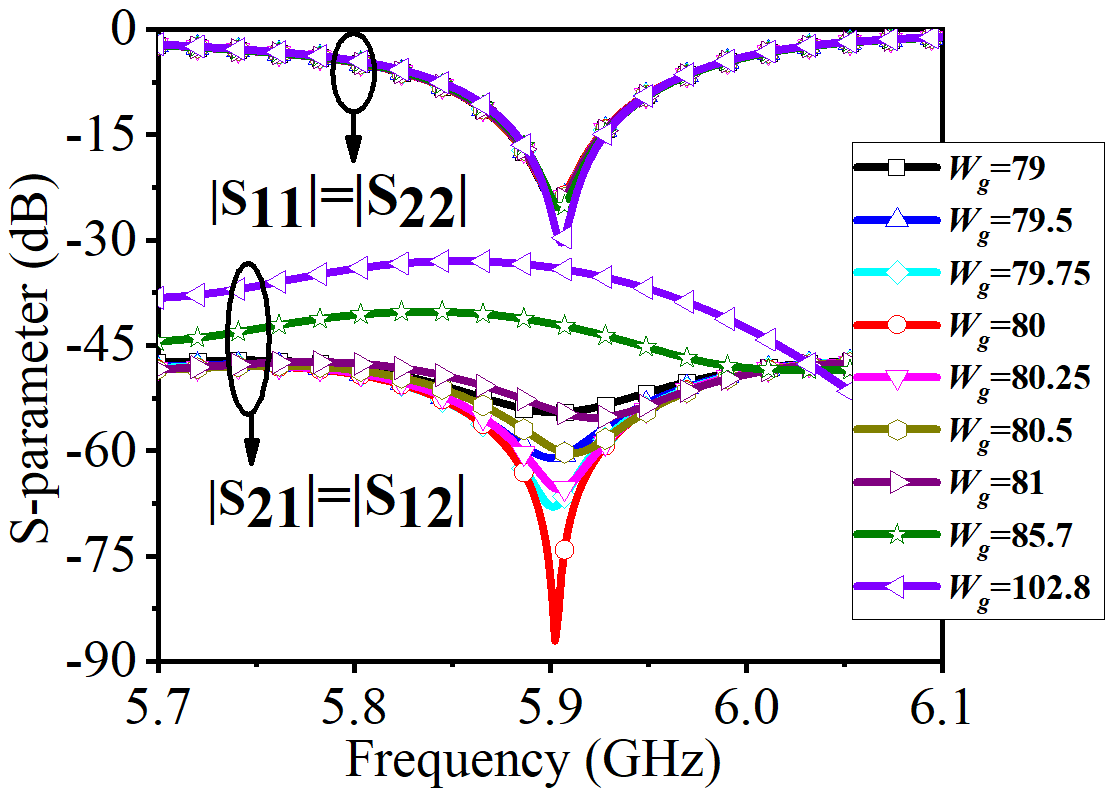}}%
\caption{Parametric studies on ground plane dimensions (a) $L_g$, and (b) $W_g$ (Unit: mm).}%
\label{fig:example}%
\end{figure}

\section{Design of Multi-Element ($1-$Tx and $2-$Rx) Full Duplex Antenna System}
The $1-$Tx and $1-$Rx FD antenna system is extended to $1-$Tx and $2-$Rx FD system to verify the design scalability of the proposed \textit{p}-SIC technique. Moreover, the multi-antenna receiver system provides the additional degree of freedom to improve the system diversity, i.e. the capability of multiple-input multiple-output (MIMO) configuration, in a multi-path crowded application environment as in ITS. The schematic of the proposed $3-$element ($1-$Tx and $2-$Rx) FD antenna system is shown in Fig. 14. The S-parameter response of this FD configuration is provided in Fig. 15, illustrates more than $70$ dB \textit{p}-SIC between $Tx_1$-$Rx_1$ and $Tx_1$-$Rx_2$ at the 5.9 GHz operating frequency, which satisfies the FD requirements. Moreover, the isolation between two Rx element is more than 30 dB over the operating band full fills the MIMO requirement. S-parameter response in Fig. 15 signifies that the proposed \textit{p}-SIC technique is also applicable to the $3-$element FD system with a minor design modification on $l_d$, $y_s$ and $g_2$. The modified design parameters are provided in Fig. 14.

 \begin{figure}[]
        \centering
         \includegraphics[width=\columnwidth,height=15 cm,keepaspectratio]{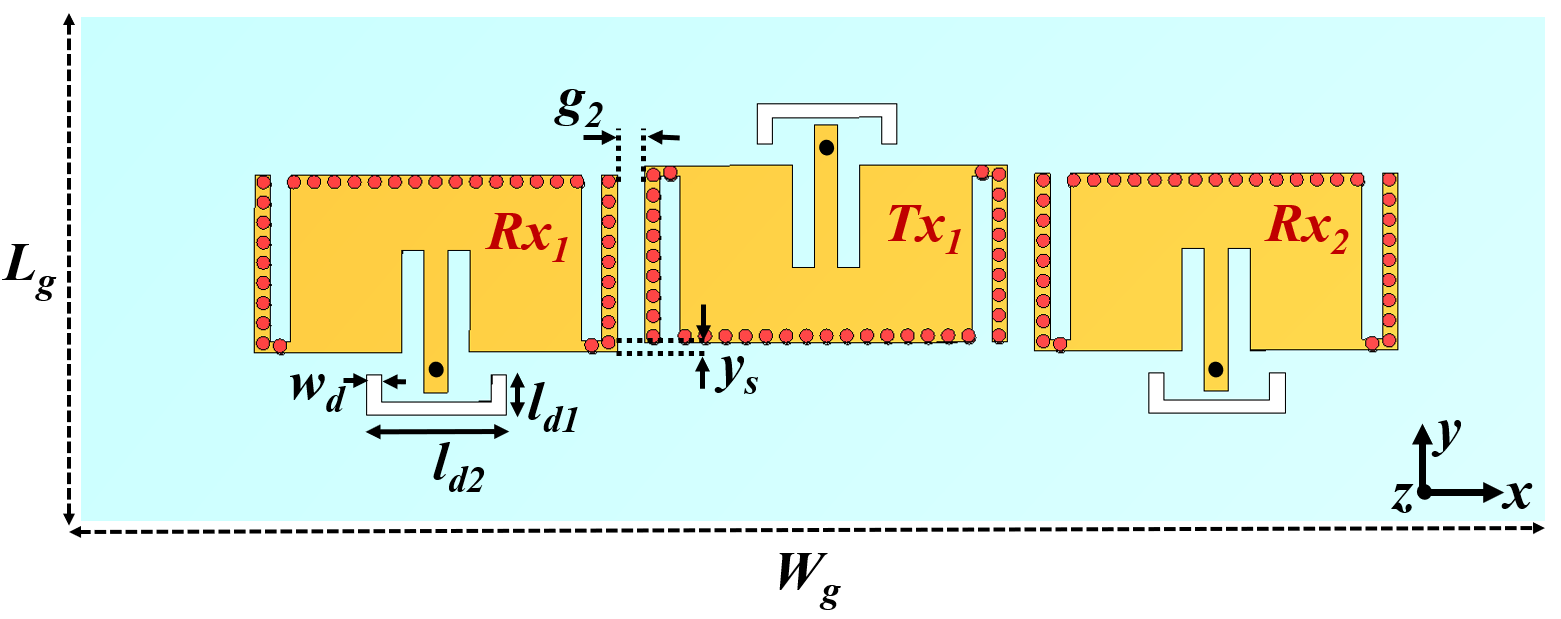}
         \caption{Schematic of 3-element (1-Tx and 2-Rx) in-set fed FD antenna system at 5.9 GHz. Dimension (in mm): $l_d = 2l_{d1}+ l_{d2} = 17.15, y_s = 0.65, g_2 = 2.$}
         \label{fig3}
    \end{figure}
    
    \begin{figure}[]
        \centering
         \includegraphics[width=\columnwidth,height=15 cm,keepaspectratio]{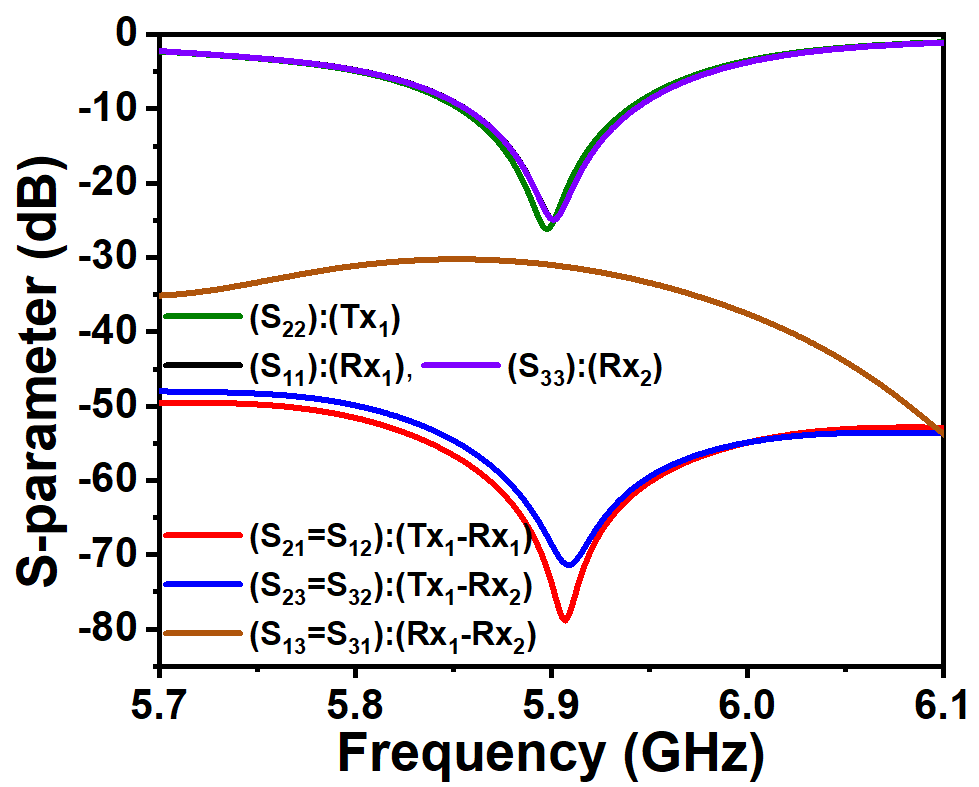}
         \caption{Frequency variation of S-parameters of 3-element (1-Tx and 2-Rx) FD antenna system as shown in Fig. 14.}
         \label{fig3}
    \end{figure}

\section{Prototype Fabrication and Measurement Results}
The fabricated prototype of the proposed two and three element FD antenna system are shown in Fig. 16 and 17 respectively. The S-parameters of these two and three element prototypes are measured using Agilent N5230A PNA and compared with the simulation results as shown in Fig. 18 and 19 respectively with their measurement set-up. Similarly , the 2D radiation pattern of Tx antenna element showing the simulated and measured curves of two and element configuration are provided in Fig. 20 and 21 respectively. The X-pol level of the two and three element configuration is below $20$ dB. The measured results are in good agreement with the simulated ones, which validates the proposed \textit{p}-SIC technique. A complete performance parameters of the proposed FD antenna configuration provided in table II.   

\begin{figure}[]
        \centering
         \includegraphics[width=\columnwidth,height=15 cm,keepaspectratio]{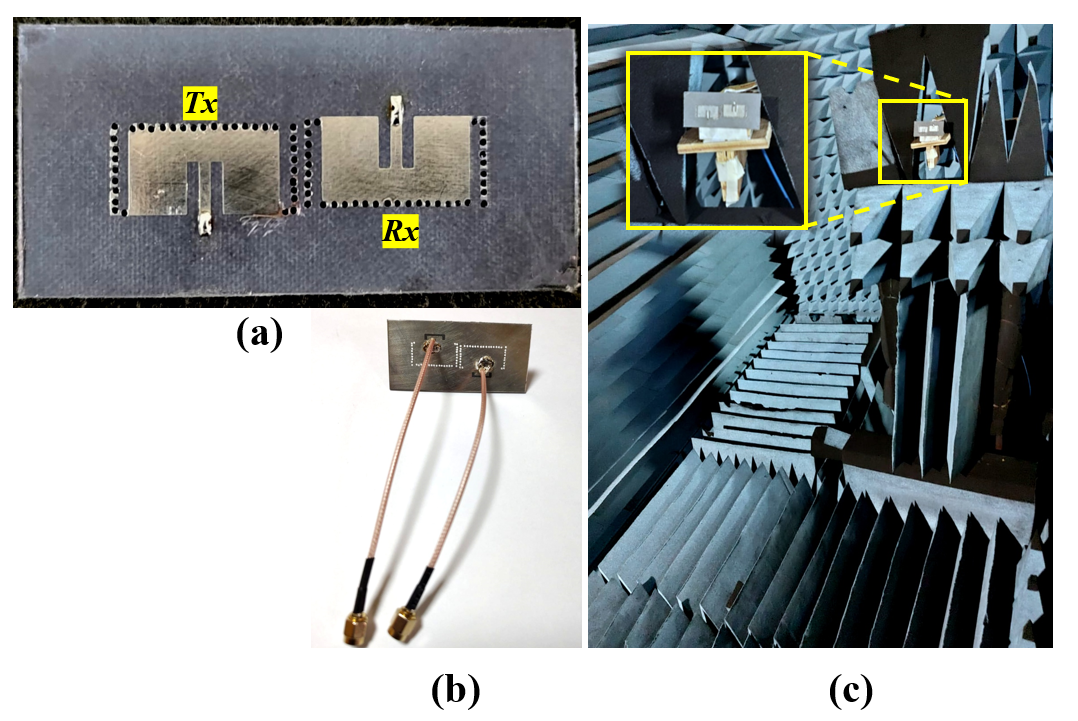}
         \caption{Fabricated prototype of the proposed 2-element FD antenna system of Fig. 3: (a) Top view, (b) Bottom view and (c) Far-field radiation pattern measurement setup in Anechoic chamber.}
         \label{fig3}
    \end{figure}
    
     \begin{figure}[]
        \centering
         \includegraphics[width=\columnwidth,height=15 cm,keepaspectratio]{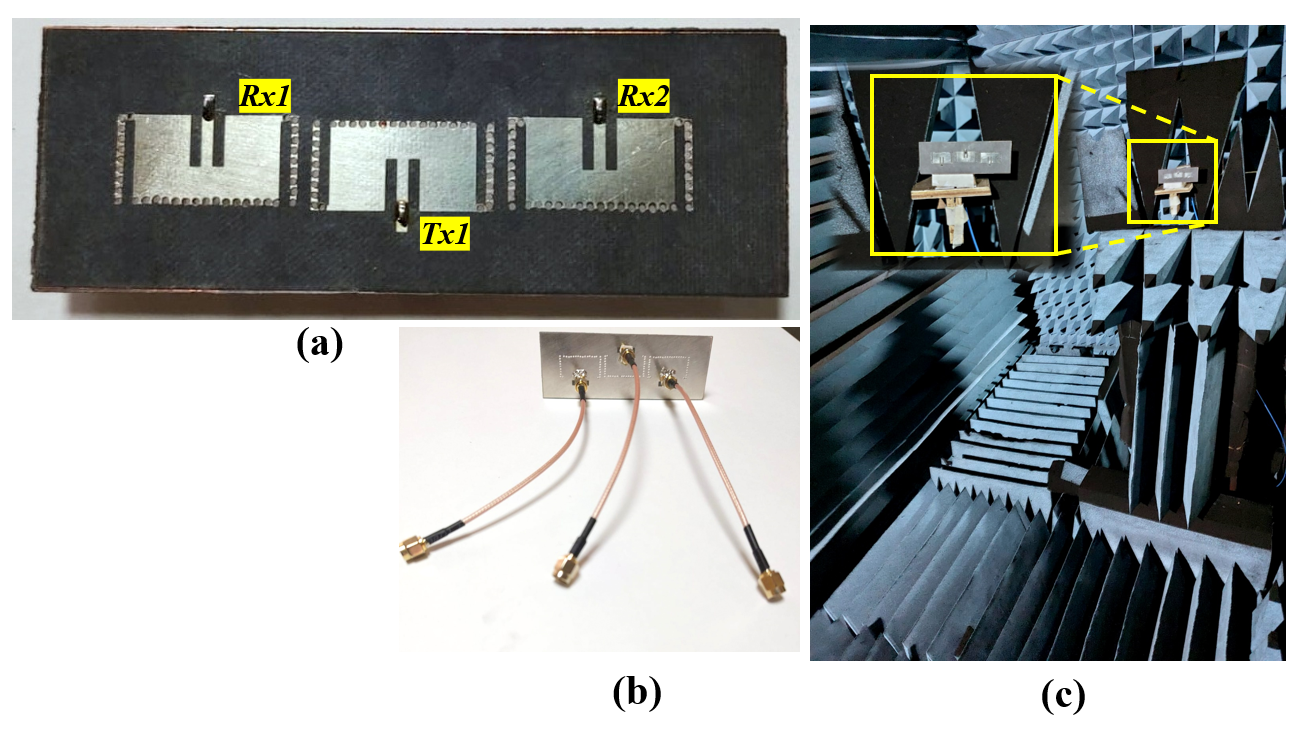}
         \caption{Fabricated prototype of the proposed 3-element FD antenna system of Fig. 14: (a) Top view, (b) Bottom view and (c) Far-field radiation pattern measurement setup in Anechoic chamber.}
         \label{fig3}
    \end{figure}

    \begin{figure}[]
        \centering
         \includegraphics[width=\columnwidth,height=8 cm,keepaspectratio]{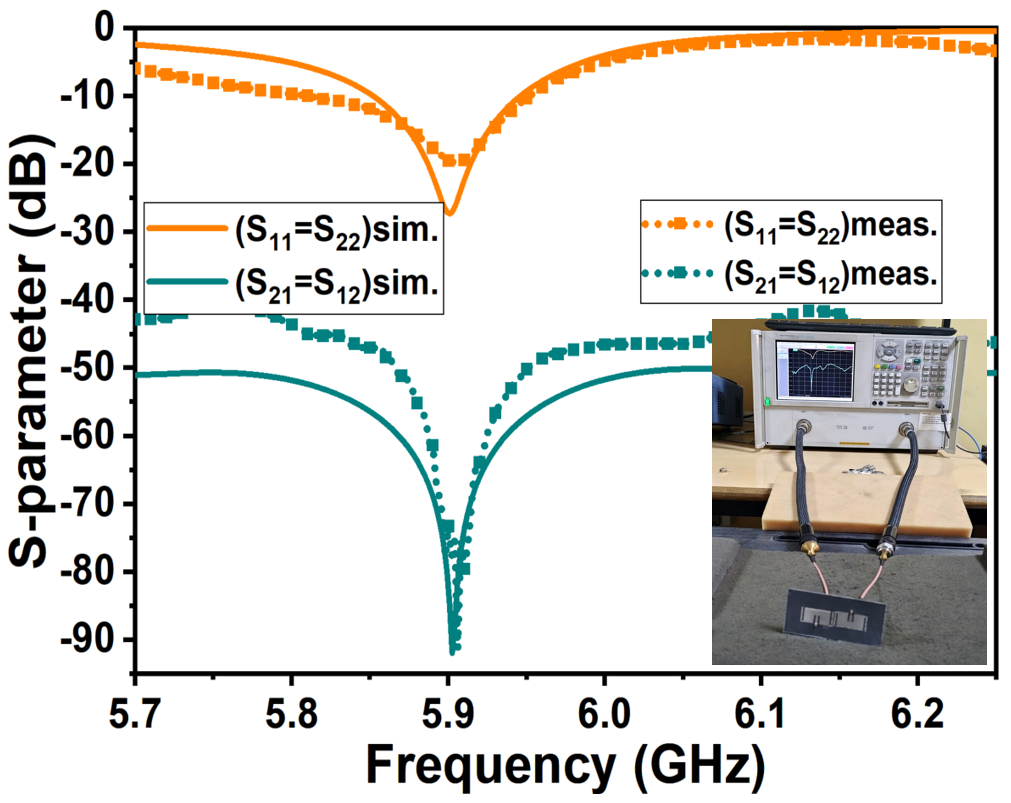}
         \caption{Frequency variation of simulate and measured S-parameters of 2-element (1-Tx and 2-Rx) FD antenna system as shown in Fig. 3.}
         \label{fig3}
    \end{figure}
    
    \begin{figure}[]
        \centering
         \includegraphics[width=\columnwidth,height=15 cm,keepaspectratio]{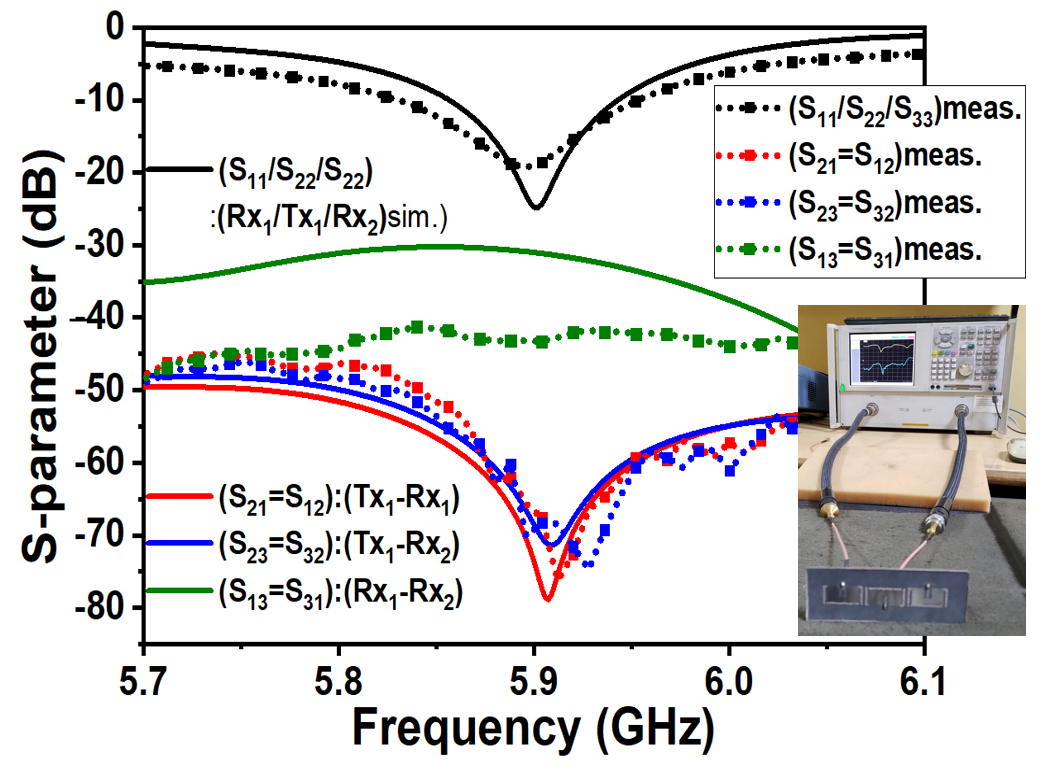}
         \caption{Frequency variation of simulate and measured S-parameters of 3-element (1-Tx and 2-Rx) FD antenna system as shown in Fig. 3.}
         \label{fig3}
    \end{figure}
    
     \begin{figure}[]
        \centering
         \includegraphics[width=\columnwidth,height=15 cm,keepaspectratio]{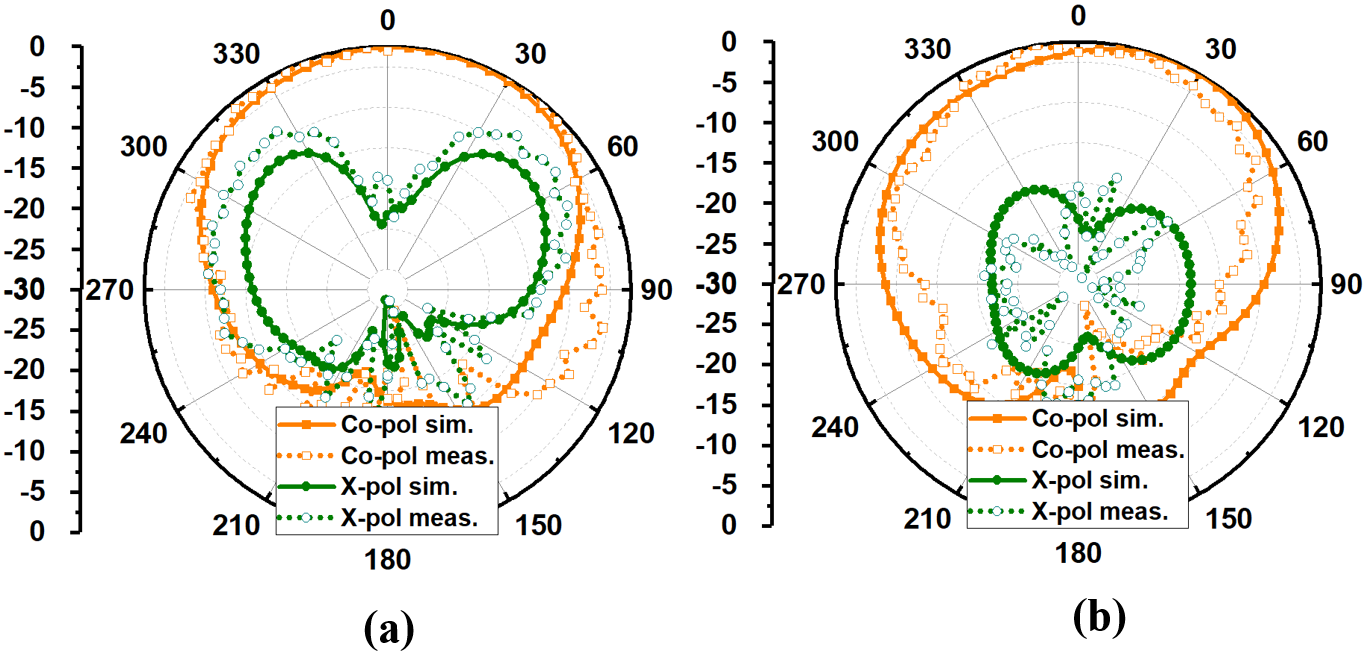}
         \caption{2D radiation pattern of the Tx antenna of 2-element prototype of Fig. 16 (a) xoz-plane, (b) yoz-plane.}
         \label{fig3}
    \end{figure}
    
    \begin{figure}[]
        \centering
         \includegraphics[width=\columnwidth,height=15 cm,keepaspectratio]{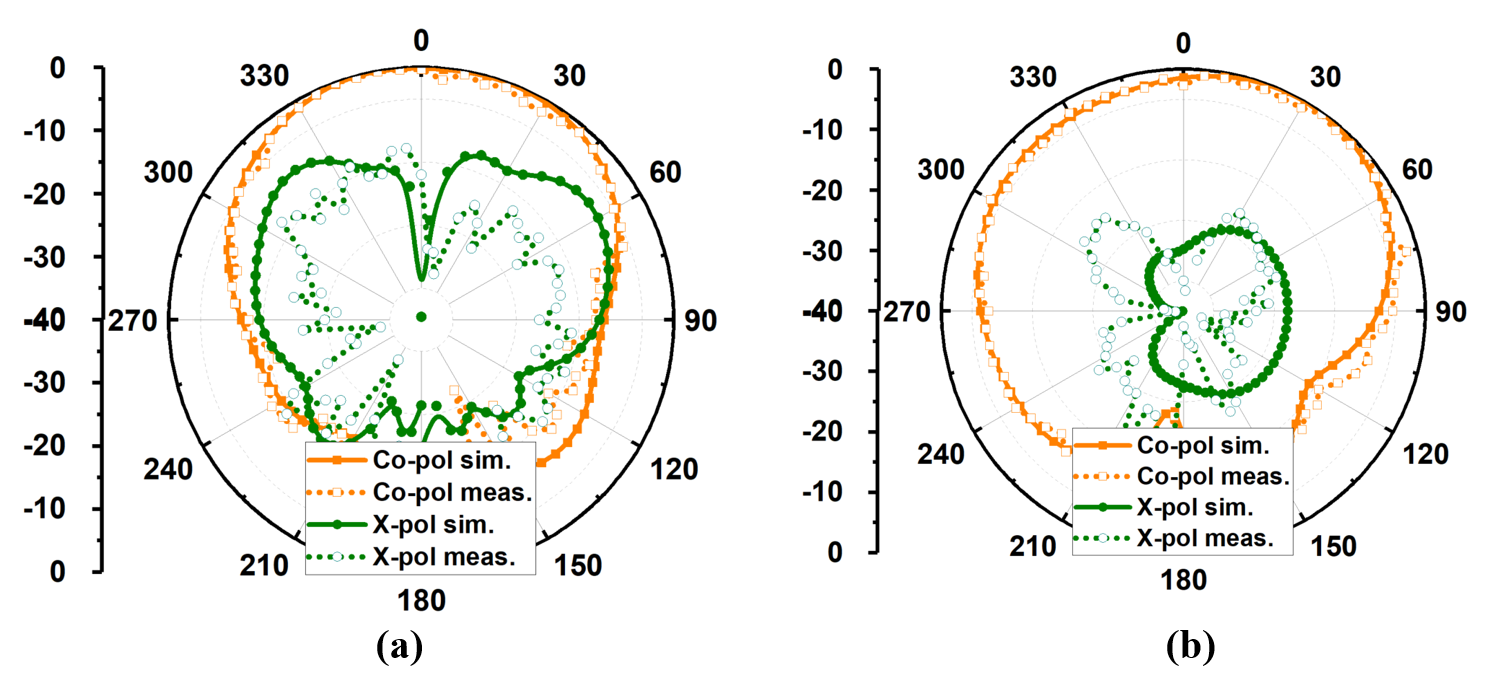}
         \caption{2D radiation pattern of the Tx antenna of 3-element prototype of Fig. 17 (a) xoz-plane, (b) yoz-plane.}
         \label{fig3}
    \end{figure}

\begin{table}[]
\centering
\caption{Proposed 2-element FD Antenna Performance Parameter}
\begin{tabular}{cc}
\hline
\textbf{Parameter} & \textbf{value} \\ \hline
No of antenna element & $2$($1$-Tx and $1$-Rx) \\
Operating Frequency & $5.9$ GHz \\
Impedance Bandwidth & $5850-5944$ MHz \\
Maximum Gain & $5.63$ dBi \\
Total efficiency & $90.5\%$ \\
Cross polar level & $< 20$ dB \\
Maximum \textbackslash{}textit\{p\}-SIC & $\approx 90$ dB \\
Application & FD ITS \\ \hline
\end{tabular}
\end{table}    
    
          \begin{figure}[]
        \centering
         \includegraphics[width=\columnwidth,height=10 cm,keepaspectratio]{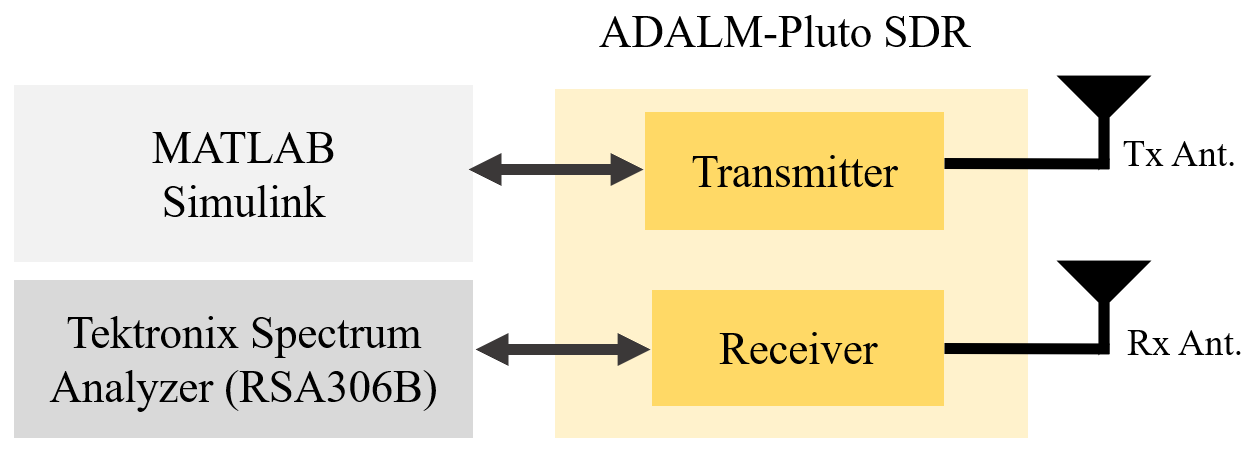}
         \caption{Schematic representation of real-time measurement setup to observe self-interference channel.}
         \label{fig3}
    \end{figure}

\subsection {Experimental Demonstration of Self-Interference Channel Between Tx and Rx and Its Suppression using Proposed \textit{p}-SIC Technique}
We carried out two-set of experiments using stage-I (See Fig. 4) and stage-IV (see Fig. 10) FD antenna to visualize the real-time self-interference between Tx and Rx. The schematic representation of experimental set-up is shown in Fig. 22, as described in \cite{hzo}. Here, we use Adalm-Pluto software defined radio (SDR) through the MATLAB's-Simulink as a source (transmitter) and Tektronix RSA306B real-time spectrum analyser as receiver connected to Tx and Rx of FD-antenna respectively as shown in Fig. 22. Fig. 23 (a) and (b) show the real-time experimental set-up using proposed stage-I and stage-IV FD-antenna along with the DPX (Digital Phosphor Technology)-spectrum indicating the self-interference between the Tx and Rx of FD antenna when operating simultaneously. It is clearly observed that the FD-antenna having least Tx-Rx coupling (see Fig. 23(a)) shows nearly no interference above system noise floor (here $-90$ dB) compared to FD-antenna having high Tx-Rx coupling (see Fig. 23(a)), which exhibits very high (nearly $40$ dB above system noise floor) signal interference. This signifies the practical utility of the proposed  \textit{p}-SIC technique.

    \begin{figure}[]
\centering

\subfloat[]{
	\label{subfig:correct}
	\includegraphics[width=0.5\textwidth]{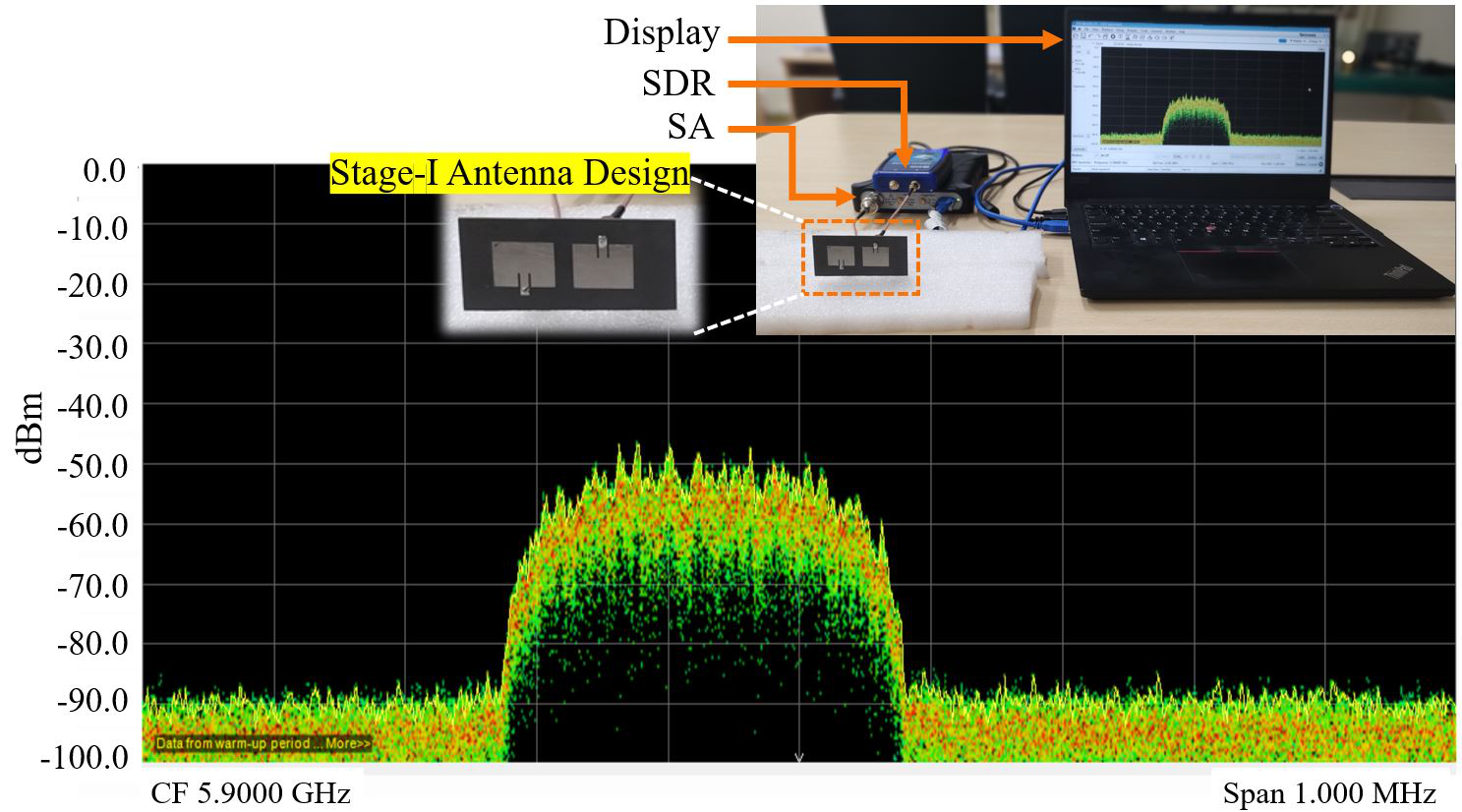} } 

\subfloat[]{
	\label{subfig:notwhitelight}
	\includegraphics[width=0.5\textwidth]{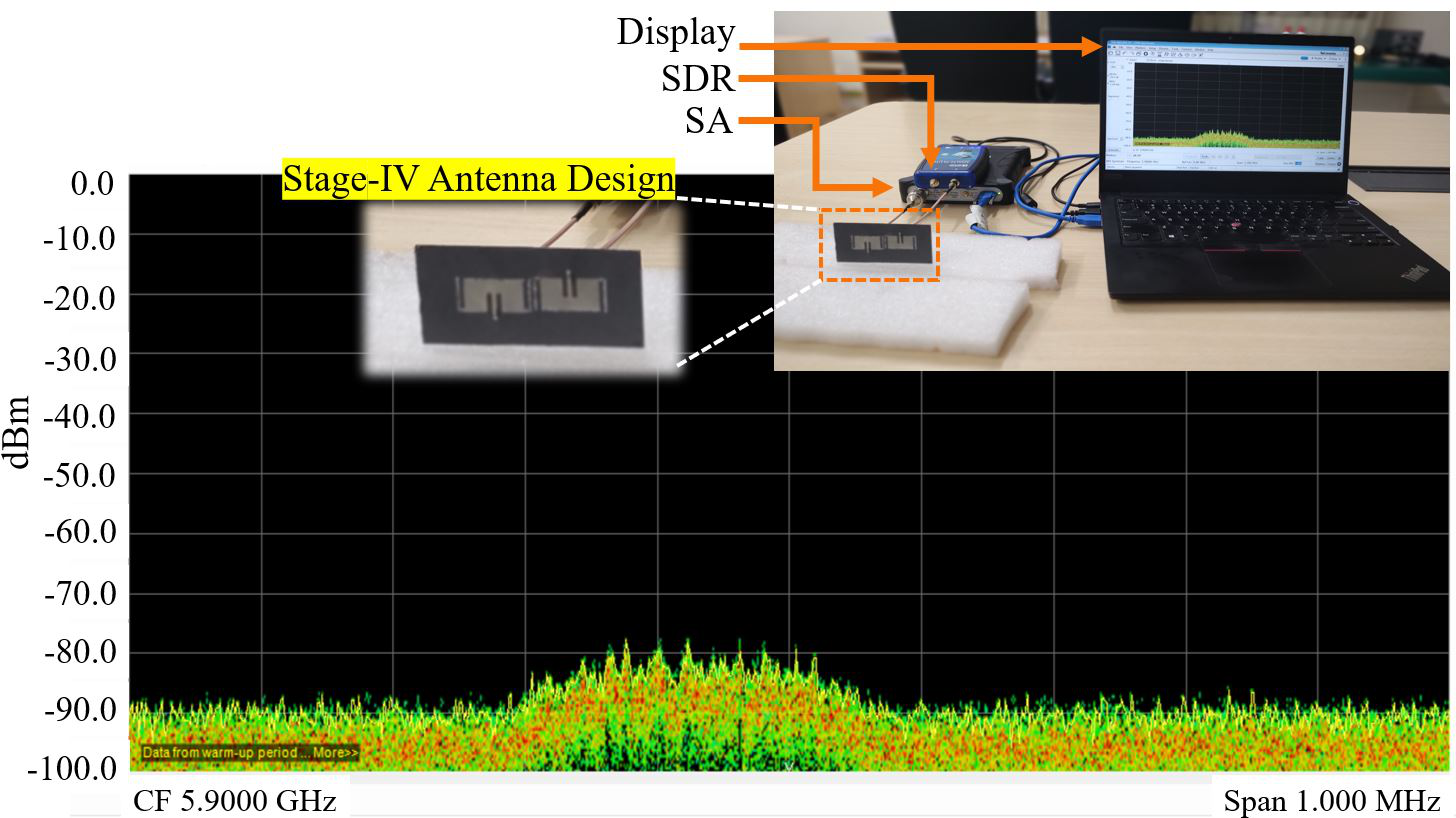} }

\caption{Received DPX (Digital Phosphor Technology)-spectrum at $5.9$ GHz opearting frequency and measurement setup (inset) showing the signal interference between the Tx and Rx of (a) Stage-I, and (b) Stage-IV FD-antenna module. (SDR: Software Defined Radio, SA: Spectrum Analyser)}
\label{kohler}

\end{figure}

\subsection {MIMO Performance Parameters}
MIMO performance study of the proposed two and three element FD antenna configuration is carried out by evaluating envelop correlation coefficient (\textit{ECC}) and channel capacity loss (\textit{CCL}) in the operating band (see Fig. 24 (a) and (b)) as discussed in \cite{jc}. The value of \textit{ECC} is much less than 0.5 in the uniform propagation environment, which indicates excellent diversity performance, which is further reflected in the low value of channel capacity loss ($< 0.5$ bits/s/Hz) in the operating band.      
 
\begin{figure}[]
\centering

\subfloat[]{
	\label{subfig:correct}
	\includegraphics[width=0.45\textwidth]{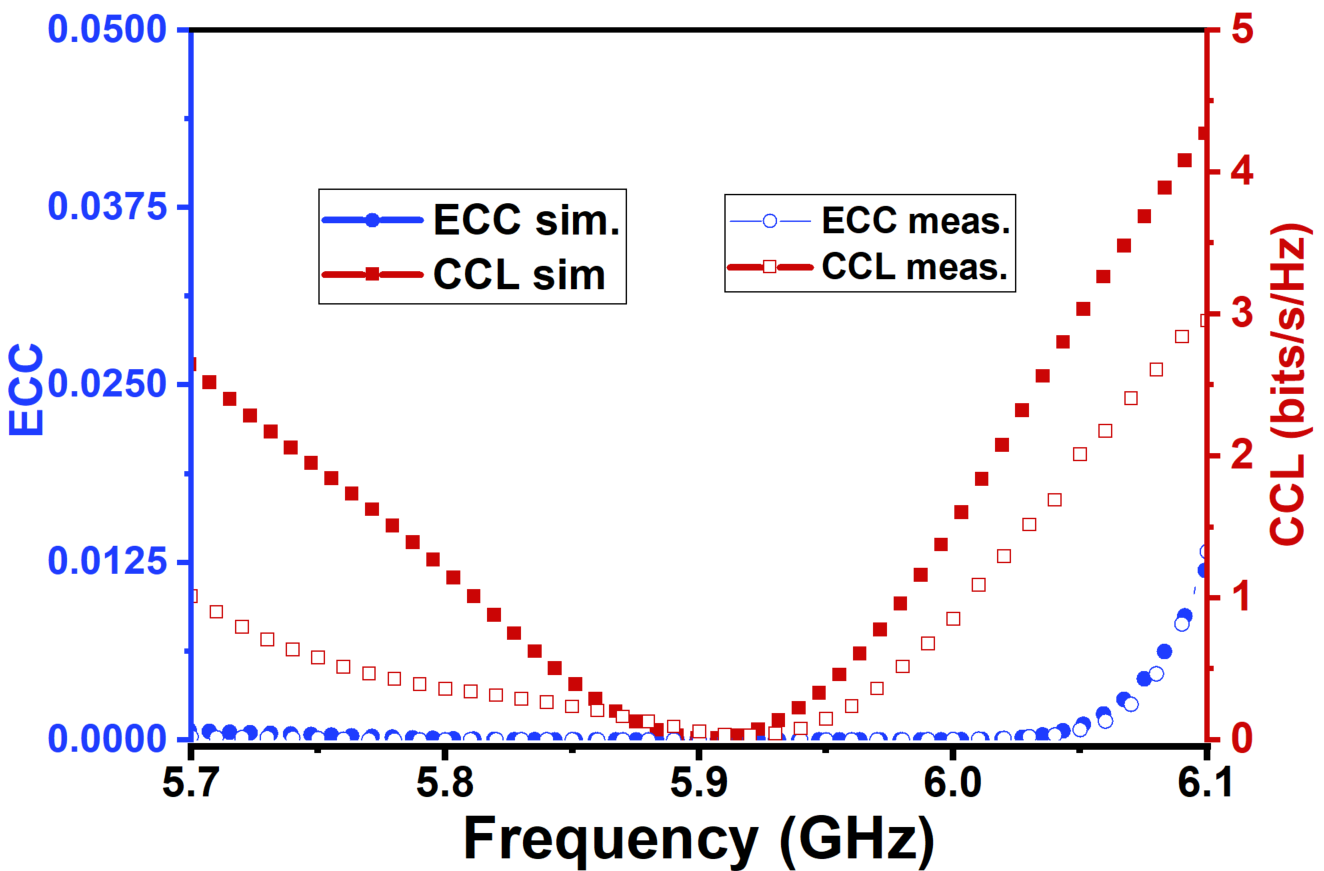} } 

\subfloat[]{
	\label{subfig:notwhitelight}
	\includegraphics[width=0.45\textwidth]{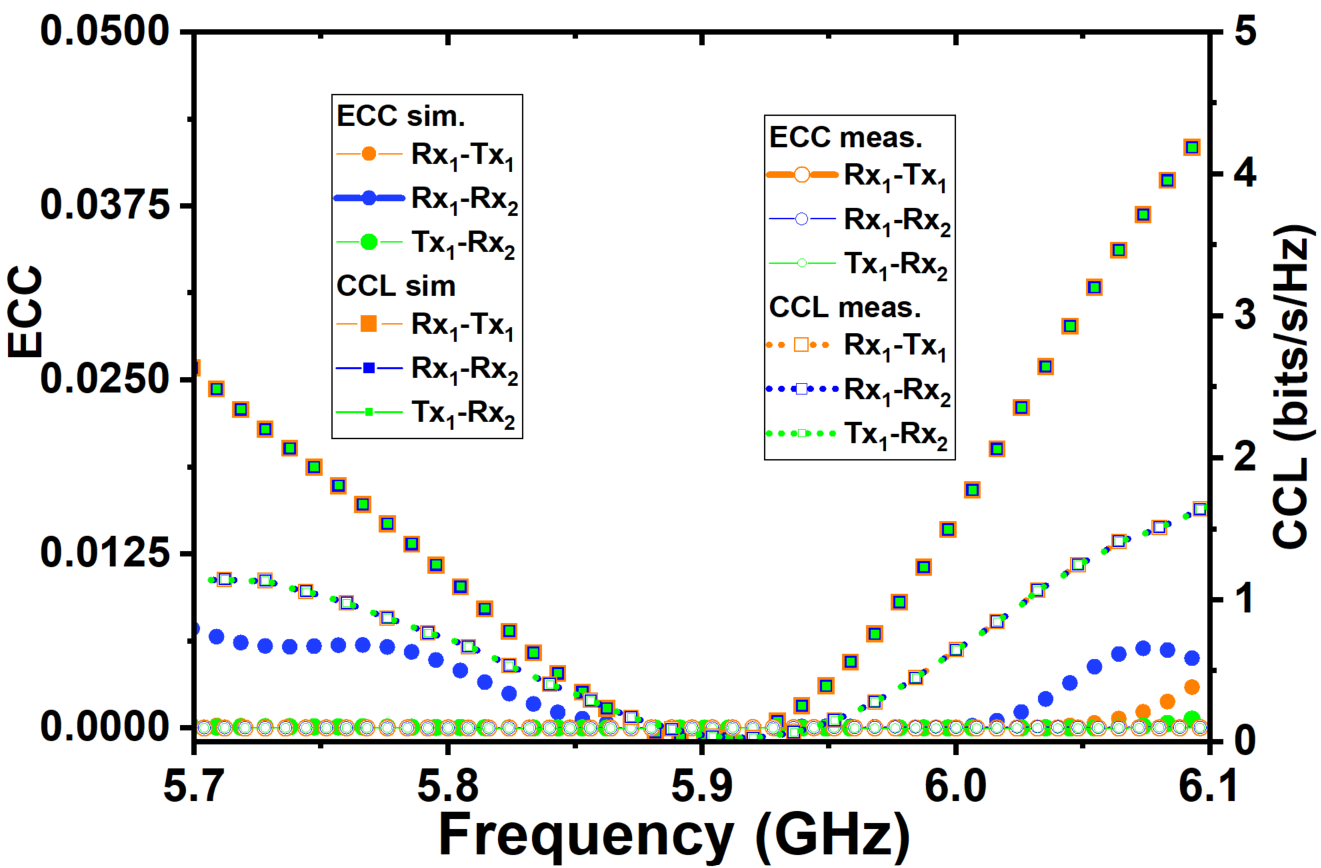} } 

	
\caption{Parametric study of design parameters: (a) $l_{d}$, (b) $y_{s}$ and (c) $g_{2}$ to observe the effect of parameter variation on FD antenna S-parameter. (Unit: mm)}
\label{kohler}

\end{figure}

\section{Conclusion}
In this paper, the design of a two-element, as well as a three-element inset-fed antenna configuration for FD implementation, are presented along with a novel \textit{p}-SIC technique between the antenna elements. The complete proposed FD antenna configuration, together with \textit{p}-SIC technique, is planar in form, easy to implement, and avoids the conventional use of bulky hybrid coupler-based FD antenna technique to achieve high \textit{p}-SIC. The mechanism of the proposed \textit{p}-SIC technique along with antenna design steps is described using field distribution between the antenna elements and DGS loaded microstrip line analogy. The \textit{p}-SIC technique provides the maximum isolation of $\approx90$ dB and $\approx80$ dB for two and element FD antenna configuration respectively. It is verified using a real-time experimental demonstration using SDR and SA that the proposed \textit{p}-SIC effectively suppresses the unwanted signal interference in Rx from its own Tx. Further, MIMO performance parameters such as ECC and CCL are evaluated to study the diversity performance of the proposed FD antenna systems. The antenna configurations are best suitable for FD ITS application.   

\section{Acknowledgement}
Authors would like to thank Prof. K. J. Vinoy, Chair, Dept. of ECE, IISc Bangalore for providing the measurement facilities (PNA and Anechoic chamber). Authors would also like to thank Mr. Debaprasad Barad (M. Tech, Res) and Mr. Yugesh Chandrakapure (M. Tech), ECE, IISc Bangalore for their help in fabrication and measurement respectively. This work is jointly supported by IoE-IISc postdoctoral fellowship and SID-IISc.

\end{document}